\documentclass[aps,prd,preprint,eqsecnum,nofootinbib,groupedaddress,floatfix]{revtex4}

\usepackage{graphicx,epsf,color,amsmath}

\def\fig#1{Fig.~{\ref{#1}}}
\def\Fig#1{Fig.~{\ref{#1}}}
\def\eqn#1{Eq.~(\ref{#1})}
\def\EqnR#1{Eq.~{\eqref{#1}}}

\def\sect#1{Sect.~{\ref{#1}}}
\def\Sect#1{Sect.~{\ref{#1}}}

\def\eps{\epsilon}
\def\pol{\varepsilon}

\def\spa#1.#2{\left\langle#1\,#2\right\rangle}
\def\spb#1.#2{\left[#1\,#2\right]}
\def\q{q}  % soft leg
\def\S{\Sigma}

\newif\ifdraft
\drafttrue
\newif\ifpreprint
\def\Ord{{\cal O}}
\newbox\charbox
\newbox\slabox
\def\s#1{{      % Feynman slash
        \setbox\charbox=\hbox{$#1$}
        \setbox\slabox=\hbox{$/$}
        \dimen\charbox=\ht\slabox
        \advance\dimen\charbox by -\dp\slabox
        \advance\dimen\charbox by -\ht\charbox
        \advance\dimen\charbox by \dp\charbox
        \divide\dimen\charbox by 2
        \raise-\dimen\charbox\hbox to \wd\charbox{\hss/\hss}
        \llap{$#1$} }}

\begin{document}

\preprint{UCLA/14/TEP/104,  $\null\hskip 9.5cm \null$
 NORDITA-2014-78}

\title{
\ifpreprint
\hbox{\rm\small
UCLA/11/TEP/??? $\null\hskip 8.cm \null$
Nordita-2014-????\break}
\hbox{$\null$\break}
\fi
Low-Energy Behavior of Gluons and Gravitons\\ from Gauge Invariance}

\author{Zvi Bern${}^a$, Scott Davies${}^a$, Paolo Di Vecchia${}^b$${}^c$ and Josh Nohle${}^a$
\\
$\null$
\\
\it 
${}^a$Department of Physics and Astronomy, UCLA, Los Angeles, CA
90095-1547, USA \\
${}^b$The Niels Bohr Institute, University of Copenhagen, Blegdamsvej 17, DK 2100 Copenhagen, Denmark \\
${}^c$Nordita, KTH Royal Institute of Technology and Stockholm University,
Roslagstullsbacken 23, SE-10691 Stockholm, Sweden \\
$\null$\\
$\null$\\
}

\begin{abstract}
We show that at tree level, on-shell gauge invariance can be used to
fully determine the first subleading soft-gluon behavior and the first two
subleading soft-graviton behaviors.  Our proofs of the behaviors for
$n$-gluon and $n$-graviton tree amplitudes are valid in $D$ dimensions
and are similar to Low's proof of universality of the first subleading
behavior of photons.  In contrast to photons coupling to massive
particles, in four dimensions the soft behaviors of gluons and
gravitons are corrected by loop effects.  We comment on how such
corrections arise from this perspective.  We also show that loop
corrections in graviton amplitudes arising from scalar loops appear
only at the second soft subleading order.  This case is particularly
transparent because it is not entangled with graviton infrared
singularities.  Our result suggests that if we set aside the issue of
infrared singularities, soft-graviton Ward identities of extended BMS
symmetry are not anomalous through the first subleading order.
\end{abstract}

\maketitle

\section{Introduction}

Interest in the soft behavior of gravitons and gluons has recently
been renewed by a proposal from Strominger and
collaborators~\cite{Strominger,CachazoStrominger} showing that
soft-graviton behavior follows from Ward identities of extended Bondi,
van der Burg, Metzner and Sachs (BMS) symmetry~\cite{BMS,
  ExtendedBMS}.  This has stimulated a variety of studies of the
subleading soft behavior of gravitons and gluons. In four spacetime
dimensions, Cachazo and Strominger~\cite{CachazoStrominger} showed
that tree-level graviton amplitudes have a universal behavior through
second subleading order in the soft-graviton momentum.  In
Ref.~\cite{SoftGluonProof} an analogous description of tree-level soft
behavior for gluons at first subleading order was given.
Interestingly, these universal behaviors hold in $D$ dimensions as
well~\cite{Volovich}. In four dimensions, there is an interesting
connection between the subleading soft behavior in gauge theory and
conformal invariance~\cite{Conformal,IntegrandSoft}.  There are also
recent constructions of twistor-related theories with the desired soft
properties~\cite{TwistorSoft}.  Soft behavior in string theory and
for higher-dimension operators has also been
discussed~\cite{StringSoft,IntegrandSoft}.

Soft theorems have a long history and were recognized in the 1950s and
1960s to be an important consequence of local on-shell gauge
invariance~\cite{LowFourPt, LowTheorem, Weinberg, OtherSoftPhotons}.
(For a discussion of the low-energy theorem for photons see Chapter 3
of Ref.~\cite{DFFR}.) For photons, Low's
theorem~\cite{LowTheorem} determines the amplitudes with a soft
photon from the corresponding amplitudes without a photon, through
$\Ord(q^0)$, where $q$ is the soft-photon momentum.  

The universal leading soft-graviton behavior was first discussed by
Weinberg~\cite{Weinberg}.  The leading behavior is uncorrected to all
loop orders~\cite{OneLoopMHVGrav}.  Using dispersion relations, Gross
and Jackiw analyzed the particular example of Compton scattering of
gravitons on massive scalar particles~\cite{GrossJackiw}.  They showed
that, for fixed angle, the Born contributions have no corrections
up to, but not including, fourth order in the soft momentum.
Jackiw then applied gauge-invariance arguments
similar to those of Low to reanalyze this case~\cite{Jackiw}.
However, for our purposes this case is too special because the
degenerate kinematics of $2\rightarrow2$ scattering leads to extra
suppression not only at tree level, but at loop level as well.
In particular, the soft limits are finite at fixed angle.
This may be contrasted with the behavior for larger numbers of legs,
where the amplitudes at all loop orders
diverge as a graviton becomes soft, matching the tree behavior.
Thus, the
results of Refs.~\cite{GrossJackiw,Jackiw} cannot be directly applied
to our discussion of $n$-point behavior.  A more recent discussion of
the generic subleading behavior of soft gluons and gravitons is given
in Refs.~\cite{WhiteYM,WhiteGrav}.

Soft-gluon and graviton behaviors are, in general, modified by loop
effects~\cite{LoopCorrections, HeHuang}. This is not surprising given
that loop corrections arising from infrared singularities occur in QCD,
starting with the leading
behavior~\cite{OneLoopSoftBern,OneLoopSoftKosower}.  We note that
Ref.~\cite{FreddyNew} proposed that by keeping the
dimensional-regularization parameter $\eps=(4-D)/2 <0$ finite as one
takes the soft limit, loop corrections can be avoided, as explicitly
shown in five-point ${\cal N}=4$ super-Yang-Mills examples.  However,
this prescription is not physically sensible because it does not get
soft physics correct and, in particular, ruins the cancellation of
leading infrared divergences in QCD. One can instead view this as a
prescription on integrands prior to loop integration; in this way, the
five-point ${\cal N} = 4$ super-Yang-Mills results in
Ref.~\cite{FreddyNew} were extended to all numbers of loops and legs for planar
amplitudes~\cite{IntegrandSoft}.

Extended BMS symmetry gives us a remarkable new understanding for the
behavior of soft gravitons in four spacetime
dimensions~\cite{Strominger}.  However, given that universal soft
behavior holds also in $D$ dimensions as well as for gluons, we expect
that there is a more general explanation not tied to four dimensions.
In this paper, we show that, just as for photons~\cite{LowTheorem},
on-shell gauge invariance can be used to fully determine subleading
behavior.  We show that in nonabelian gauge theory, on-shell
gauge invariance dictates that at tree level the first subleading term
is universal and controlled by the amplitude with the soft gluon
removed. Similarly, in gravity the first two subleading terms at tree
level are universal.  Our proof is valid in $D$ dimensions
because it uses only on-shell gauge invariance together with 
$D$-dimensional three-point vertices.

We shall also explain how loop corrections arise in this context.
In nonabelian gauge theory and gravity, there are
``factorizing'' loop corrections to the three-vertex controlling the
soft behavior.  However, in gravity, generically the dimensionful nature
of the coupling implies that there are no loop corrections to the
leading behavior~\cite{OneLoopMHVGrav}, no corrections beyond one loop
to the first subleading behavior, and no corrections beyond two loops
to the second subleading behavior~\cite{LoopCorrections}.

As shown long ago, in gauge theory the factorizing contributions are
suppressed: In gauge theory they vanish at leading order in the soft
limit~\cite{OneLoopSoftBern,OneLoopSoftKosower}, but are nontrivial at
the first subleading order~\cite{LoopCorrections,HeHuang}.  Similarly,
we prove that for the case of a scalar circulating in the loop, the
factorizing loop corrections to the soft-graviton behavior vanish not
only for the leading order but for the first subleading order as well.
This case is particularly transparent because there are no infrared
singularities~\cite{GravityIR} or contributions to the soft operators
arising from them.  We expect that for all other particles circulating
in the loop, only contributions associated with infrared singularities
will appear at the first subleading soft order.  Indeed, this
suppression has been observed in the explicit examples of
infrared-finite amplitudes studied in
Refs.~\cite{LoopCorrections,HeHuang}.  These results suggest that, up
to issues associated with infrared singularites, the soft Ward
identities of BMS symmetry~\cite{Strominger} are not anomalous.  We
note that while there are loop corrections to the first subleading
soft-graviton behavior linked with infrared singularities, they come
from a well-understood source and therefore should not be too
disruptive when studying the connection to BMS symmetry.

This paper is organized as follows. In \Sect{PhotonSection}, we review
Low's theorem for the case of a soft photon coupled to $n$ scalars,
showing how gauge invariance determines the first subleading behavior.
In \Sect{GravitonSection}, we repeat the analysis for a soft graviton.
Next, in \Sect{GluonSection}, we study the case of a soft gluon where
all external particles are gluons and discuss spin contributions in
some detail. The analysis for a soft graviton is extended to the case
where all external particles are gravitons in
\Sect{PureGravitonSection}.  In \Sect{LoopSection}, we explain how loop
corrections to the soft operators arise from the perspective of
on-shell gauge invariance and show that there are no corrections to
the first subleading soft-graviton behavior for scalars in the loop.
We give our conclusions in \Sect{ConclusionSection}.

\subsection*{Added note}

While this manuscript was being finalized, a paper appeared 
constraining soft behavior using Poincar{\' e} and gauge invariance,
as well as from a condition arising from the
distributional nature of scattering amplitudes~\cite{NewPaper}.  In
this way, the authors determine the form of the subleading soft
differential operators up to a numerical constant for every
leg.

%%%%%%%%%%%%%%%%%%%%%%%%%%%%%%%%%%%%%%%%%%%%%%%
\section{Photon soft limit with $n$ scalar particles}
\label{PhotonSection}

In this section, we review the classic theorem due to
Low~\cite{LowTheorem} on the subleading soft behavior of photons, for
simplicity focusing on the case of a single photon coupled to $n$
scalars. As explained by Low in 1958, gauge invariance enforces the
universality of the first subleading behavior, allowing us to
fully determine it in terms of the amplitude without the soft photon.  In
subsequent sections, we will apply a similar analysis to cases with
gravitons and gluons.

%%%%%%%%% FIGURE %%%%%%%%%%%%%%%                                              
\begin{figure}[tb]
\begin{center}
\vskip .7 cm 
\includegraphics[scale=.4]{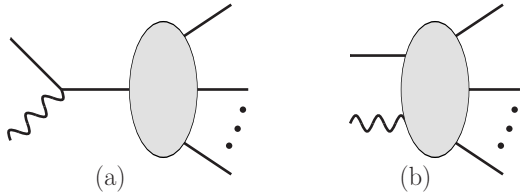}
\end{center}
\vskip -.7 cm 
\caption[a]{\small  Diagrams of the form (a) give universal leading soft behavior. 
The subleading behavior comes from both diagrams types (a) and (b). 
\label{TreeSoftFigure}
}
\end{figure}
%%%%%%%%%%%%%%%%%%%%%%%%%%%%%%%%

As illustrated in \fig{TreeSoftFigure}, the scattering amplitude of a
single photon and $n$ scalar particles arises from (a) contributions 
with a pole in the soft momentum $\q$ and (b) contributions with no pole:
\begin{align}
A_n^{\mu} (\q; k_1, \ldots, k_{n} ) = \sum_{i=1}^{n} e_i \frac{k_i^{\mu}}{k_i\cdot \q}
  T_n(k_1, \ldots, k_i+ \q, \ldots, k_n) + N_n^{\mu} (\q; k_1, \ldots, k_n ) \,.
\label{MTN}
\end{align}
For our purposes, it is convenient to not include the polarization vectors until the end of
the discussion.  The full amplitude is obtained by contracting
$A_n^{\mu}$ with the physical photon polarization $\pol_{q\mu}$. 
The first term in \eqn{MTN} corresponds to the
emission of the photon from one of the scalar external lines as
illustrated in \fig{TreeSoftFigure}(a) and is divergent in the
soft-photon limit, while the second term, illustrated in
\fig{TreeSoftFigure}(b), is finite in the soft-photon limit.  The
electric charge of particle $i$ is $e_i$.

On-shell gauge invariance implies 
\begin{align}
0 &= \q_{\mu} A_n^{\mu}(\q; k_1, \dots, k_n ) \notag \\
&= \sum_{i=1}^{n} e_i  T_n(k_1, \ldots, k_i + \q,  \ldots,  k_n) +
q_{\mu} N_n^{\mu} (\q; k_1, \dots, k_n )\,,
\label{gaugeinv}
\end{align}
valid for any value of $\q$. 
Expanding around $\q=0$, we have 
\begin{align}
0 &= \sum_{i=1}^{n} e_i \, \left[T_n(k_1, \ldots, k_i,  \ldots,  k_n) +\q_\mu \frac{\partial}{\partial k_{i\mu}} 
           T_n(k_1, \ldots, k_i,  \ldots,  k_n)\right] \nonumber \\
& \hspace{.5cm} 
   \vphantom{\frac{1}{1}}\hspace{,5cm}+ q_{\mu} N_n^{\mu} (\q=0; k_1, \dots, k_n )+\Ord(q^2)\,.
\label{ExpandedPhoton}
\end{align}
At leading order, this equation is
\begin{align}
\sum_{i=1}^n e_i  = 0\,,
\end{align}
which is simply a statement of charge conservation~\cite{Weinberg}.
At the next order, we have
\begin{align}
\q_{\mu} N_n^{\mu} (0; k_1, \ldots, k_n) = 
- \sum_{i=1}^n e_i \q_{\mu} \frac{\partial}{\partial k_{i\mu}} T_n(k_1, \dots, k_n) \,.
\label{subleading}
\end{align}
This equation tells us that $N_n^{\mu}(0; k_1, \ldots, k_n)$ is
entirely determined up to potential pieces that are separately gauge
invariant. However, it is easy to see that the only expressions local
in $q$ that vanish under the gauge-invariance condition
$q_{\mu}E^{\mu}=0$ are of the form,
\begin{align}
E^{\mu}=(B_1\cdot\q)B_2^{\mu}-(B_2\cdot\q)B_1^{\mu} \,,
\label{GaugeInvariantPhotonTerms}
\end{align}
where $B_1^{\mu}$ and $B_2^{\mu}$ are arbitrary vectors that are local
in $q$ and constructed with the momenta of the scalar particles.  The
explicit factor of the soft momentum $\q$ in each term means that they
are suppressed in the soft limit and do not contribute to
$N_n^{\mu}(0; k_1, \ldots, k_n)$.  We can therefore remove the $\q_{\mu}$ from
\EqnR{subleading}, leaving
\begin{align}
 N_n^{\mu} (0; k_1, \ldots, k_n) = 
- \sum_{i=1}^n e_i \frac{\partial}{\partial k_{i\mu}} T_n (k_1, \dots, k_n)\,,
\end{align}
thereby determining $N_n^{\mu} (0; k_1, \ldots, k_n)$ 
as a function of the amplitude without the photon.
Inserting this into \EqnR{MTN} yields
\begin{align}
A_n^{\mu} (q; k_1, \dots, k_n ) 
= \sum_{i=1}^{n} \frac{e_i}{k_i\cdot\q} 
 \left[ k_i^{\mu}  -i q_{\nu} J_i^{\mu \nu} \right]  T_n (k_1, \dots, k_n) 
+ \Ord(q)\,,
\label{Oq}
\end{align}
where
\begin{align}
J_i^{\mu \nu}\equiv i\left(k_i^{\mu} \frac{\partial}{\partial k_{i\nu}} -  k_i^{\nu} \frac{\partial}{\partial k_{i\mu}}\right) \,,
\label{OrbitalOperator}
\end{align}
is the orbital angular-momentum operator
and $T_n(k_1, \dots, k_n) $ is the scattering amplitude involving $n$ scalar particles. 
\EqnR{Oq} is  Low's theorem for the case of one photon and
$n$ scalars.  

Low's theorem is unchanged at loop level for the simple reason that even at
loop level, all diagrams containing a pole in the soft momentum are of
the form shown in \fig{TreeSoftFigure}(a), with loops appearing only
in the blob and not correcting the external vertex.  If the scalars
are massive, the integrals will not have infrared
discontinuities that could lead to loop corrections 
of the type described in Ref.~\cite{LoopCorrections}.

It is also interesting to see if there is any further information at higher
orders in the soft expansion.  If we go one order further in the
expansion, we find the extra condition,
\begin{align}
\frac{1}{2} \sum_{i=1}^{n} e_i q_{\mu} q_{\nu} 
\frac{ \partial^{2}}{ \partial k_{i\mu} \partial k_{i\nu} }T_n (k_1, \dots, k_n) +q_{\mu} q_{\nu}  \frac{\partial N_n^{\mu}}{\partial q_{\nu}} (0; k_1, \dots, k_n) =0\,.
\label{2ndorder}
\end{align}
This implies
\begin{align}
 \sum_{i=1}^{n} e_i 
\frac{\partial^{2}}{\partial k_{i\mu} \partial k_{i\nu} }T_n (k_1, \dots, k_n)
 + \left[\frac{\partial N_n^{\mu}}{\partial q_{\nu}}
+ \frac{\partial N_n^{\nu}}{\partial q_{\mu}}\right] (0; k_1, \dots, k_n)=0\,,
\label{2ndord}
\end{align}
where the final set of arguments belongs to both terms in the bracket.
Gauge invariance determines only the symmetric
part of the quantity $ \frac{\partial N_n^{\nu}}{\partial q_{\mu}} (0;
k_1, \dots, k_n)$. The antisymmetric part is not fixed by gauge
invariance; indeed, this corresponds exactly to terms of the type in
\eqn{GaugeInvariantPhotonTerms}.  Then, up to this order, we have
\begin{align}
A_n^{\mu} (q; k_1, \dots, k_n)&= \sum_{i=1}^{n} \frac{e_i}{k_i\cdot\q} \left[  k_i^{\mu}  -i q_{\nu} J_i^{\mu \nu} \left( 1+ \frac{1}{2}
 q_{\rho} \frac{\partial}{\partial k_{i \rho}} \right) \right]  T_n (k_1, \dots, k_n)  \nonumber \\
 &\hspace{.5cm}+
\frac{1}{2} q_{\nu} \left[  \frac{\partial N_n^{\mu}}{\partial q_{\nu}} - \frac{\partial N_n^{\nu}}{\partial q_{\mu}}\right] (0; k_1, \dots, k_n)
+O(q^2)\,.
\label{3or}
\end{align}
It is straightforward to see that one gets zero by saturating the
previous expression with $q_{\mu}$.

In order to write our universal expression in terms of the amplitude,
we contract $A_n^{\mu}(\q; k_1, \dots, k_n )$ with the photon
polarization $\pol_{q\mu}$.  From \eqn{Oq}, we have the soft-photon limit
of the single-photon, $n$-scalar amplitude:
\begin{align}
A_n(q;k_1,\ldots,k_n)\rightarrow\left[S^{(0)}+S^{(1)}\right]T_n(k_1,\ldots,k_n)+\Ord(q)\,,
\end{align}
where
\begin{align}
S^{(0)}&\equiv\sum_{i=1}^ne_i\frac{k_i\cdot\pol_q}{k_i\cdot\q}\,, \notag \\
S^{(1)}&\equiv-i\sum_{i=1}^ne_i\frac{\pol_{q\mu}q_{\nu}J_i^{\mu\nu}}{k_i\cdot\q}\,,
\end{align}
and $J_i^{\mu\nu}$ is given in \eqn{OrbitalOperator}.

%%%%%%%%%%%%%%%%%%%%%%%%%%%%%%%%%%%%%%%%%%%
\section{Graviton soft limit with $n$ scalar particles}
\label{GravitonSection}

We now turn to the case of gravitons coupled to $n$ scalars.  We shall
see that in the graviton case, gauge invariance can be used to
fully determine the first two subleading orders in the soft-graviton
momentum $q$.  Together with the subsequent sections, this shows that
the tree behavior through second subleading soft order uncovered in
Ref.~\cite{CachazoStrominger} can be understood as a consequence of
on-shell gauge invariance.

In the case of a graviton scattering on $n$ scalar particles, 
Eq.~(\ref{MTN}) becomes
\begin{align}
M_n^{\mu \nu } (q; k_1, \dots ,k_n ) = 
\sum_{i=1}^{n} \frac{ k_i^{\mu} k_i^{\nu} }{k_i\cdot\q} T_n (k_1, \dots, k_i+q, \dots, k_n) +
N_n^{\mu \nu} (q; k_1, \dots, k_n )\,,
\label{MTNgra}
\end{align}
where $N_n^{\mu \nu}(\q; k_1, \dots, k_n )$ is symmetric under the
exchange of $\mu$ and $\nu$.  For simplicity, we have set the
gravitational coupling constant to unity.  Similar to the gauge-theory
case, we contract with the graviton polarization tensor
$\pol_{q\mu\nu}$ at the end.  On-shell gauge invariance of the soft
leg requires that the amplitude be invariant under the shift in the
polarization tensor,
\begin{equation}
\pol_{q \mu\nu} \rightarrow \pol_{q \mu\nu} + q_\mu \pol_{q \nu} f(q,k_i)\,,
\label{GravPolShift}
\end{equation}
where $\pol_{q \nu}$ satisfies $\pol_{q \nu}\cdot q = 0$ and $f(q, k_i)$
is an arbitrary function of the momenta.   This implies that
\begin{align}
0&=q_{\mu} M_n^{\mu \nu }(\q; k_1, \dots, k_n ) \notag \\
&= \sum_{i=1}^{n} k_i^{\nu} T_n  (k_1, \dots, k_i+q, \dots, k_n) +
q_{\mu} N_n^{\mu \nu} (q; k_1, \dots, k_n )\,.
\label{genecova}
\end{align}
Strictly speaking, Eq.~\eqref{genecova} is true only after contracting
the $\nu$ index with either $\varepsilon_{q\nu}$ or a conserved current.  
Since we contract with polarizations at the end, we can use 
\eqn{genecova}.  At leading
order in $q$, we then have
\begin{align}
\sum_{i=1}^{n} k_i^{\mu} =0\,,
\label{momcon}
\end{align}
which is satisfied due to momentum conservation. (As noted by
Weinberg~\cite{Weinberg}, had there been different couplings to the
different particles, it would have prevented this from vanishing in
general; this shows that gravitons have universal coupling.)

At first order in $q$,  \eqn{genecova} implies
 \begin{align}
\sum_{i=1}^{n} k_i^{\nu} \frac{\partial }{\partial k_{i \mu}} T_n (k_1, \dots, k_n )+ N_n^{\mu \nu} (0 ; k_1, \dots, k_n ) =0\,,
\label{1ordq}
\end{align}
while at second order in $q$, it gives
\begin{align}
\sum_{i=1}^{n} k_i^{\nu} \frac{\partial^2}{\partial k_{i \mu} \partial k_{i \rho}} T_n(k_1, \dots, k_n)
+\left[\frac{\partial N_n^{\mu \nu} }{\partial q_{\rho}}+ \frac{\partial N_n^{\rho \nu} }{\partial q_{\mu}}\right](0 ; k_1, \dots, k_n )=0\,.
\label{2ndorfe}
\end{align}
As in the case of the photon, this is true up to gauge-invariant
contributions to $N_n^{\mu\nu}$.  However, the requirement of locality
prevents us from writing any expression that is local in $q$ yet not
sufficiently suppressed in $q$.  In fact, the most general local
expression that obeys the gauge-invariance condition $q_\mu E^{\mu\nu}
= q_\nu E^{\mu\nu} = 0$ is of the form,
\begin{align}
E^{\mu\nu} &=\Big((B_1\cdot\q)B_2^{\mu}-(B_2\cdot\q)B_1^{\mu}\Big)\Big((B_3\cdot\q)B_4^{\nu}-(B_4\cdot\q)B_3^{\nu}\Big)\,,
\label{GravityGaugeInvariant}
\end{align}
where the $B_i^{\mu}$ are local in $q$ and constructed in terms of the
momenta of the scalar particles.  In the amplitude, $E^{\mu\nu}$ will
be contracted against the symmetric traceless graviton-polarization
tensor $\varepsilon_{q\mu\nu}$, so there is no need to include
potential terms proportional to $q^\mu$, $\q^\nu$ or $\eta^{\mu \nu}$.
The two powers of $q$ in \eqn{GravityGaugeInvariant} mean that such
terms do not contribute to the soft expansion at the orders in which
we are interested.

Using Eqs.~\eqref{1ordq} and \eqref{2ndorfe} in Eq.~\eqref{MTNgra}, we write the expression for a soft graviton as
\begin{align}
M_n^{\mu \nu } (q; k_1 \dots k_n ) &= \sum_{i=1}^{n} \frac{k_i^{\nu}}{k_i\cdot\q} \left[  k_i^{\mu}  -i q_{\rho} J_i^{\mu \rho} \left( 1 + \frac{1}{2}  q_{\sigma} \frac{\partial}{\partial k_{i\sigma}} \right) \right]  T_n (k_1, \dots, k_n)  \nonumber \\
&\hspace{.5cm}+  \frac{1}{2}q_{\rho} \left[ \frac{\partial N_n^{\mu \nu}}{\partial q_{\rho}} - \frac{\partial N_n^{\rho \nu}}{\partial q_{\mu}} \right](0 ; k_1, \dots, k_n ) +\mathcal{O}(q^2)\,.
\label{gravit}
\end{align}
This is essentially the same as \eqn{3or} for the photon except 
that there is a second Lorentz index in the graviton case. We will show that, unlike the
case of the photon, the antisymmetric quantity in the second line of
the previous equation can also be determined from the amplitude
$T_n(k_1, \dots, k_n )$ without the graviton.

But, before we proceed further, let us check gauge
invariance. Saturating the previous expression with $q_{\mu}$, we see
that the first term is vanishing because of momentum conservation,
while all other terms are vanishing because $q_{\mu} q_{\rho}$ is
saturated with terms that are antisymmetric in $\mu$ and $\rho$. If,
instead, we saturate the amplitude with $q_{\nu}$, the first term is
vanishing as before due to momentum conservation, while the first term
depending on angular momentum is vanishing because of angular-momentum
conservation. The remaining terms are
\begin{align}
q_{\nu} M_n^{\mu \nu}(q ; k_1, \dots, k_n ) &= \frac{1}{2} q_{\rho} q_{\sigma}\bigg\{\sum_{i=1}^{n} \left( k_i^{\mu} \frac{\partial }{\partial k_{i\rho}} - k_i^{\rho} \frac{\partial }{\partial k_{i\mu}} \right)\frac{\partial}{\partial k_{i\sigma}}T_n(k_1, \dots, k_n ) \notag \\
&\hspace{4cm} +\left[ \frac{\partial N_n^{\mu \sigma}}{\partial q_{\rho}}  -  \frac{\partial N_n^{\rho \sigma}}{\partial q_{\mu}}\right](0;k_1, \dots, k_n )\bigg\} \notag \\
&=0\,,
\label{gaugeinv32}
\end{align}
where the vanishing follows from Eq.~(\ref{2ndorfe}), remembering that
$N_n^{\mu \nu}$ is a symmetric matrix. Therefore the amplitude in
Eq.~(\ref{gravit}) is gauge invariant. Actually, Eq.~(\ref{2ndorfe})
allows us to write the relation\,,
\begin{align}
-i\sum_{i=1}^{n} J_i^{\mu \rho}\frac{\partial}{\partial k_{i\sigma}}T_n(k_1, \dots, k_n )=\left[\frac{\partial N_n^{\rho \sigma}}{\partial q_{\mu}} -  \frac{\partial N_n^{\mu \sigma}}{\partial q_{\rho}}\right](0;k_1, \dots, k_n )\,,
\label{ideyt}
\end{align}
which fixes the antisymmetric part of the derivative of $N_n^{\mu
  \nu}$ in terms of the amplitude $T_n(k_1, \dots, k_n )$ without the
graviton.  Inserting this into Eq.~(\ref{gravit}), we can then rewrite
the terms of $\Ord(q)$ as follows:
\begin{align}
M_n^{\mu \nu}(q ; k_1, \dots, k_n )\big|_{\Ord(q)}&= -\frac{i}{2} \sum_{i=1}^{n}  \frac{q_{\rho} q_{\sigma}}{k_i\cdot\q} \left[
k_i^{\nu} J_i^{\mu \rho} \frac{\partial}{\partial k_{i\sigma}} - k_i^{\sigma} J_i^{\mu \rho} \frac{\partial}{\partial k_{i\nu}}   \right]T_n(k_1, \dots, k_n ) \nonumber \\
&=  -\frac{i}{2} \sum_{i=1}^{n}  \frac{q_{\rho} q_{\sigma}}{k_i\cdot\q} \left[ J_i^{\mu \rho}k_i^{\nu}\frac{\partial}{\partial k_{i\sigma}}-  \left( J_i^{\mu \rho} k_{i\nu} \right)     \frac{\partial}{\partial k_{i\sigma}}  \right. \nonumber \\
&\hspace{2.7cm} \left. - J_i^{\mu \rho}k_i^{\sigma} \frac{\partial}{\partial k_{i\nu}}   +  \left( J_i^{\mu \rho} k_i^{\sigma} \right) \frac{\partial}{\partial k_{i\nu}}   \right]T_n(k_1, \dots, k_n ) \nonumber \\
&= \frac{1}{2} \sum_{i=1}^{n}  \frac{1}{k_i\cdot\q} \left[\Big((k_i\cdot q)( \eta^{\mu \nu} q^{\sigma} -  q^{\mu} \eta^{\nu\sigma} ) - k_i^{\mu} q^{\nu} q^{\sigma}\Big)\frac{\partial}{\partial k^{\sigma}_{i}}\right.  \nonumber \\
&\hspace{5cm}\left.\vphantom{\frac{\partial}{\partial}}-q_{\rho}J_i^{\mu \rho} q_{\sigma}  J_i^{\nu \sigma}\right]T_n(k_1, \dots, k_n )\,.
\label{M1}
\end{align}
Finally, we wish to write our soft-limit expression in terms of the amplitude, so we contract with the
physical polarization tensor of the soft graviton, $\pol_{q\mu\nu}$.
We see that the physical-state conditions set to zero the terms in \eqn{M1}
that are proportional to $\eta^{\mu\nu}$, $q^\mu$ and $q^\nu$.  We are then left with the following expression for the graviton soft
limit of a single-graviton, $n$-scalar amplitude:
\begin{align}
M_n(q; k_1, \dots, k_n ) \rightarrow\left[S^{(0)}+S^{(1)}+S^{(2)}\right]T_n(k_1, \dots, k_n)+\Ord(q^2)\,,
\label{Mcomplete}
\end{align}
where
\begin{align}
&S^{(0)}\equiv \sum_{i=1}^{n}\frac{\pol_{\mu \nu}k_i^{\mu} k_i^{\nu}}{k_i\cdot\q}\,, \notag \\
&S^{(1)}\equiv-i\sum_{i=1}^{n}\frac{\pol_{\mu \nu}k_i^{\mu}q_{\rho}J_i^{\nu \rho}}{k_i\cdot\q}\,, \notag \\
&S^{(2)}\equiv-\frac{1}{2}\sum_{i=1}^{n}\frac{\pol_{\mu \nu}q_{\rho} J_i^{\mu \rho} q_{\sigma}  J_i^{\nu \sigma}}{k_i\cdot\q}\,.
\end{align}
These soft factors follow from gauge invariance and agree with those computed in Ref.~\cite{CachazoStrominger}.

We have also looked at higher-order terms and found that gauge invariance 
does not fully determine them in terms of derivatives acting on 
$T_n(k_1, \dots, k_n)$.

%%%%%%%%%%%%%%%%%%%%%%%%%%%%%%%%%%%%%%%%%%%%%%%%%%%

\section{Soft limit of $n$-gluon amplitudes}
\label{GluonSection}

%%%%%%%%% FIGURE %%%%%%%%%%%%%%%                                              
\begin{figure}[tb]
\begin{center}
\vskip .7 cm 
\includegraphics[scale=.4]{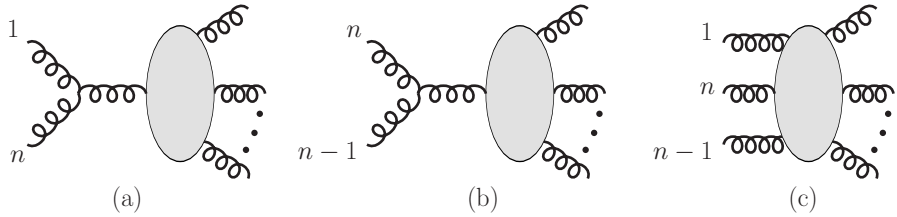}
\end{center}
\vskip -.7 cm 
\caption[a]{\small  Diagrams (a) and (b) give leading universal soft-gluon 
behavior. The first
subleading behavior of the amplitude contained in the non-pole diagram (c) 
can be determined via on-shell gauge invariance. 
\label{GluonSoftFigure}
}
\end{figure}
%%%%%%%%%%%%%%%%%%%%%%%%%%%%%%%%

\subsection{Behavior of gluon tree amplitudes}

In this section, we generalize the procedure of \sect{PhotonSection} to
the case of $n$-gluon tree amplitudes prior to turning to the case of
$n$ gravitons in the next section. As we shall discuss in
\sect{LoopSection}, the soft-gluon behavior has loop corrections.

We consider a tree-level color-ordered amplitude (see
e.g. Ref.~\cite{DixonReview}) where gluon $n$ becomes soft, where we
define $q \equiv k_n$.  As before, we find it convenient to contract
the expression with polarization vectors only at the end to obtain the
full amplitude.  In the case of $n$ gluons, we have two pole terms:
one where the soft gluon is attached to leg 1 (see
Fig.~\ref{GluonSoftFigure}(a)) and the other where the soft gluon is
attached to leg $n-1$ (see Fig.~\ref{GluonSoftFigure}(b)).  In
addition to the contributions containing a pole in the soft momentum,
we have the usual term $N_n^{\mu ; \mu_1 \cdots \mu_{n-1}} ( q; k_1,
\dots, k_{n-1}) $ that is regular in the soft limit (see
Fig.~\ref{GluonSoftFigure}(c)).  Together, the contributions in
\fig{GluonSoftFigure} give
\begin{align}
& A_n^{\mu;\mu_1 \cdots \mu_{n-1}} (q; k_1, \dots, k_{n-1}) 
\notag \\
&\hspace{1cm}=
\frac{\delta^{\mu_{1}}_{\rho}k_1^{\mu}
+\eta^{\mu \mu_1} q_{\rho}
-\delta^{\mu}_{\rho} q^{\mu_1}}
{\sqrt{2}(k_1\cdot\q)}
\,A_{n-1}^{\rho\mu_2 \cdots \mu_{n-1}}  (k_1 +q, k_2,  \dots, k_{n-1})
\nonumber \\
&\hspace{1.5cm}
-\frac{\delta^{\mu_{n-1}}_{\rho} k_{n-1}^{\mu}
+\eta^{\mu_{n-1} \mu}  q_{\rho}
-\delta^{\mu}_{\rho} q^{\mu_{n-1}}}
{\sqrt{2}(k_{n-1}\cdot\q)}
\,A^{\mu_1 \cdots \mu_{n-2}\rho}_{n-1} (k_1 , \dots, k_{n-2}, k_{n-1}+q )
\nonumber \\
&\hspace{1.5cm}
+N_n^{\mu ; \mu_1 \cdots \mu_{n-1}} ( q; k_1, \dots, k_{n-1})\,.
\label{totampli0}
\end{align}
We have dropped terms from the three-gluon vertex that
vanish when saturated with the external-gluon polarization vectors in
addition to using the current-conservation conditions,
\begin{align}
&(k_1+q)_{\rho}\,A_{n-1}^{\rho \mu_2 \cdots \mu_{n-1}}  (k_1 +q, k_2,  \dots, k_{n-1})=0\,, \notag \\
&(k_{n-1}+q)_{\rho}\,A^{\mu_1 \cdots \mu_{n-2}\rho}_{n-1} (k_1 , \dots, k_{n-2}, k_{n-1}+q )=0\,,
\end{align}
which are valid once we contract with the polarization vectors carrying 
the $\mu_j$ indices.  By introducing the spin-one
angular-momentum operator,
\begin{align}
(\Sigma_i^{\mu\sigma})^{\mu_{i}\rho} \equiv
i\left(\eta^{\mu\mu_{i}}\eta^{\rho\sigma}-\eta^{\mu\rho}\eta^{\mu_{i}\sigma}\right)\,,
\label{SpinAngularMomOperator}
\end{align}
we can write \eqn{totampli0} as
\begin{align}
& A_n^{\mu;\mu_1 \cdots \mu_{n-1}} (q; k_1, \dots, k_{n-1}) 
\notag \\
&\hspace{1cm}=
\frac{\delta^{\mu_{1}}_{\rho}k_1^{\mu}
-iq_{\sigma}(\Sigma_{1}^{\mu\sigma})^{\mu_{1}}_{\phantom{\mu_{1}}\rho}}
{\sqrt{2}(k_1\cdot\q)}
\,A_{n-1}^{\rho\mu_2 \cdots \mu_{n-1}}  (k_1 +q, k_2,  \dots, k_{n-1})
\nonumber \\
&\hspace{1.5cm}
-\frac{\delta^{\mu_{n-1}}_{\rho} k_{n-1}^{\mu}
-iq_{\sigma}(\Sigma_{n-1}^{\mu\sigma})^{\mu_{n-1}}_{\phantom{\mu_{n-1}}\rho}}
{\sqrt{2}(k_{n-1}\cdot\q)}
\,A^{\mu_1 \cdots \mu_{n-2}\rho}_{n-1} (k_1 , \dots, k_{n-2}, k_{n-1}+q )
\nonumber \\
&\hspace{1.5cm}
+N_n^{\mu ; \mu_1 \cdots \mu_{n-1}} ( q; k_1, \dots, k_{n-1})\,.
\label{totampli}
\end{align}
Notice that the spin-one terms independently vanish when contracted with $q_{\mu}$.

The on-shell gauge invariance of Eq.~\eqref{totampli} requires
\begin{align}
0&=q_{\mu}A_n^{\mu; \mu_1 \cdots \mu_{n-1}} (q; k_1, \dots, k_{n-1}) \notag \\
&= \frac{1}{\sqrt{2}}\,A_{n-1}^{\mu_1\mu_2 \cdots \mu_{n-1}}  (k_1 +q , k_2, \dots, k_{n-1})-\frac{1}{\sqrt{2}}\,A_{n-1}^{\mu_1 \cdots \mu_{n-2}\mu_{n-1}} (k_1 , \dots, k_{n-2}, k_{n-1}+q )   \nonumber \\
&\hspace{.5cm} + q_{\mu} N_n^{\mu;\mu_1 \cdots \mu_{n-1}} (q;k_1, \dots, k_{n-1})\,.
\label{poleterm}
\end{align}
For $q=0$, this is automatically satisfied. 
At the next order in $q$, we obtain
\begin{align}
   -\frac{1}{\sqrt{2}}\left[\frac{ \partial }{ \partial {k_{1\mu}} }-   \frac{ \partial }{ \partial {k_{n-1\mu}} }\right]A_{n-1}^{\mu_1\cdots\mu_{n-1}} 
(k_1 , k_2 \dots k_{n-1} ) =   N_n^{\mu;\mu_1\cdots \mu_{n-1}} (0;k_1, \dots, k_{n-1})\,.
\label{1order1}
\end{align}
Similar to the photon case, we ignore local gauge-invariant terms in $
N_n^{\mu;\mu_1 \cdots \mu_{n-1}} $ because they are necessarily
of a higher order in $q$.  Thus,  $N_n^{\mu;\mu_1 \cdots
  \mu_{n-1}}(0; k_1, \dots, k_{n-1})$ is determined in terms of an
expression without the soft gluon. With this, the total expression
in \eqn{totampli} becomes
\begin{align}
A_n^{\mu;\mu_1 \cdots \mu_{n-1}} (q; k_1 \dots k_{n-1})&=  \left(\frac{k_1^{\mu} }{\sqrt{2}(k_1\cdot\q)} - \frac{k_{n-1}^{\mu} }{\sqrt{2}(k_{n-1}\cdot\q)}    \right) A_{n-1}^{\mu_1 \cdots \mu_{n-1}} (k_1 , \dots, k_{n-1}) \nonumber \\
&\hspace{.5cm}-i\frac{q_{\sigma} ( J_1^{\mu \sigma} )^{\mu_1}_{\phantom{\mu_{1}}\rho}}{\sqrt{2}(k_1\cdot\q)}
\,A_{n-1}^{\rho\mu_2 \cdots \mu_{n-1}}
(k_1, \dots, k_{n-1}) \notag \\
&\hspace{.5cm}+i\frac{q_{\sigma} ( J_{n-1}^{\mu \sigma} )^{\mu_{n-1}}_{\phantom{\mu_{n-1}}\rho}}{\sqrt{2}(k_{n-1}\cdot\q)}
\,A^{\mu_1 \cdots \mu_{n-2}\rho}_{n-1} (k_1, \dots, k_{n-1})   
+ \Ord(q)\,,
\label{finordq0}
\end{align}
where 
\begin{align}
(J_i^{\mu\sigma})^{\mu_{i}\rho}
\equiv
L_i^{\mu\sigma}\eta^{\mu_i\rho}
+(\Sigma_i^{\mu\sigma})^{\mu_{i}\rho},
\label{Ji}
\end{align}
the spin-one angular-momentum operator is given in \eqn{SpinAngularMomOperator}, and the orbital angular-momentum operator is 
\begin{align}
L_i^{\mu\sigma}
\equiv
i\left(k_i^{\mu} \frac{\partial}{\partial k_{i \sigma}} -
k_i^{\sigma} \frac{\partial}{\partial k_{i \mu}}\right)\,.
\label{OrbitalAngularMomOperator}
\end{align}
Both angular-momentum operators satisfy the same commutation relations,
\begin{align}
&[ L_i^{\mu \nu} , L_i^{\rho \sigma}] = i\left(\eta^{\nu \rho} L_i^{\mu \sigma} 
   + \eta^{\mu \rho} L_i^{\sigma \nu} + \eta^{\mu \sigma} L_i^{\nu \rho} 
   + \eta^{\nu \sigma} L_i^{\rho \mu}\right)\,, \nonumber \\
& [ \Sigma_i^{\mu \nu} , \Sigma_i^{\rho \sigma}] = i\left(\eta^{\nu \rho} \Sigma_i^{\mu \sigma} 
   + \eta^{\mu \rho} \Sigma_i^{\sigma \nu} + \eta^{\mu \sigma} \Sigma_i^{\nu \rho} 
   + \eta^{\nu \sigma} \Sigma_i^{\rho \mu}\right)\,,
\label{commutators}
\end{align}
where the suppressed indices on $\Sigma_i^{\mu\nu}$ should be treated as 
matrix indices.

In order to write the final result in terms of full amplitudes, we
contract with external polarization vectors.  On the right-hand side
of Eq.~\eqref{finordq0}, we must pass polarization vectors
$\pol_{1\mu_1}$ and $\pol_{n-1\mu_{n-1}}$ through the spin-one
angular-momentum operator such that they will contract with the $\rho$
index of, respectively, $A_{n-1}^{\rho\mu_2 \cdots \mu_{n-1}}(k_1,
\dots, k_{n-1})$ and $A^{\mu_1 \cdots \mu_{n-2}\rho}_{n-1} (k_1,
\dots, k_{n-1})$.  It is convenient write the spin angular-momentum 
operator as
\begin{align}
\pol_{i\mu_i}(\Sigma_i^{\mu\sigma})^{\mu_{i}}_{\hphantom{\mu_i}\rho}A^{\rho}=i\left(\pol_i^{\mu}\frac{\partial}{\partial\pol_{i\sigma}}-\pol_i^{\sigma}\frac{\partial}{\partial\pol_{i\mu}}\right)\pol_{i\rho}A^{\rho}\,.
\end{align}
We may therefore write
\begin{align}
A_n(q;k_1,\ldots,k_{n-1})\rightarrow\left[S_n^{(0)}+S_n^{(1)}\right]
A_{n-1}(k_1,\ldots,k_{n-1})+\mathcal{O}(q)\,,
\label{eq:finalSoftGluon}
\end{align}
where
\begin{align}
&S^{(0)}_{n}
\equiv
\frac{k_{1}\cdot\pol_{n}}{\sqrt{2}\,(k_{1}\cdot q)}
-\frac{k_{n-1}\cdot\pol_{n}}{\sqrt{2}\,(k_{n-1}\cdot q)}\,, \notag \\
&S^{(1)}_{n}
\equiv
-i\pol_{n\mu}q_{\sigma}\left(\frac{J_1^{\mu\sigma}}{\sqrt{2}\,(k_{1}\cdot q)}
-\frac{J_{n-1}^{\mu\sigma}}{\sqrt{2}\,(k_{n-1}\cdot q)}\right).
\label{SoftGluonOperators}
\end{align}
Here
\begin{align}
J_i^{\mu\sigma}\equiv L_i^{\mu\sigma}+\Sigma_i^{\mu\sigma}\,,
\end{align}
where
\begin{align}
\Sigma_i^{\mu\sigma}\equiv i\left(\pol_i^{\mu}\frac{\partial}{\partial\pol_{i\sigma}}-\pol_i^{\sigma}\frac{\partial}{\partial\pol_{i\mu}}\right)\,.
\label{eq:spinTermx}
\end{align}
In using \eqn{eq:finalSoftGluon}, one must interpret $L_i^{\mu\sigma}$
as not acting on explicit polarization vectors, i.e.,
$L_i^{\mu\sigma}\pol_i=0$.  If one instead interprets polarization
vectors as functions of momenta (see e.g. Sect. 5.9 of
Ref.~\cite{WeinbergBook}) and returns a nonzero value for
$L_i^{\mu\sigma}\pol_i$, then one should not include the spin term
\eqref{eq:spinTermx}.  To be concrete, the action of the total
angular-momentum operator on momenta and polarizations is given by
\begin{align}
&J_i^{\mu\sigma}k_i^{\rho}=i\left(\eta^{\sigma\rho}k_i^{\mu}-\eta^{\mu\rho}k_i^{\sigma}\right)\,, \notag \\
&J_i^{\mu\sigma}\pol_i^{\rho}=i\left(\eta^{\sigma\rho}\pol_i^{\mu}-\eta^{\mu\rho}\pol_i^{\sigma}\right)\,.
\end{align}
We comment more on the action of the operator on polarization vectors
in \sect{spinterms}.

In conclusion, the first two leading terms in the soft-gluon expansion
of an $n$-gluon amplitude are given directly in terms of the amplitude
without the soft gluon. This derivation is valid in $D$ dimensions.
We have explicitly checked the soft-gluon formula
(\ref{eq:finalSoftGluon}) using explicit four-, five- and six-gluon
tree amplitudes of gauge theory in terms of formal polarization
vectors.

%%%%%%%%%%%%%%%%%%%%%%%%

\subsection{Connection to spinor helicity}
\label{spinterms}

To connect with the spinor-helicity formalism used in
e.g. Refs.~\cite{CachazoStrominger,LoopCorrections,HeHuang}, we show
that, up to a gauge transformation, the action of the above subleading soft
operators on polarization vectors expressed in terms of
spinor helicity is identical to the ones defined as differential
operators acting on spinors.  In the spinor-helicity formalism,
polarization vectors are expressed directly in terms of spinors
depending on the momenta:
\begin{equation}
\pol_{i}^{+\,\rho}(k_{i},k_{r})
=\frac{\left<r\right|\gamma^{\rho}\left|i\right]}{\sqrt{2}\spa{r}.{i}}\,, 
\hspace{2cm}
\pol_{i}^{-\,\rho}(k_{i},k_{r})
=-\frac{\left<i\right|\gamma^{\rho}\left|r\right]}{\sqrt{2}\spb{r}.{i}}\,,
\label{PolSpinor}
\end{equation}
where $k_i$ is the momentum of gluon $i$ and $k_r$ is a null reference
momentum. Henceforth, we will leave the $k_{i}$ argument implicit and
only display the reference momentum. The spinors are standard Weyl
spinors. We follow the conventions of Ref.~\cite{DixonReview} aside
from our use of angle and square brackets instead of the $\pm$
angle-bracket convention. In our convention, we have
\begin{align}
\langle i | = \langle i^{-}|\,,
\hspace{1cm}
[i |=\langle i^{+} | \,,
\hspace{1cm}
|i\rangle = |i^{+}\rangle \,,
\hspace{1cm}
|i] = |i^{-}\rangle \,.
\end{align}

In terms of spinors, the subleading soft factor for a tree-level
gauge-theory amplitude is~\cite{SoftGluonProof}
\begin{align}
S_{n} ^{(1)\lambda}=\frac{1}{\langle(n-1)\,n\rangle}
\tilde{\lambda}_n^{\dot{\alpha}}
\frac{\partial}{\partial\tilde{\lambda}_{n-1}^{\dot{\alpha}}}
- \frac{1}{\langle 1\,n\rangle}\tilde{\lambda}_n^{\dot{\alpha}}
  \frac{\partial}{\partial{\tilde{\lambda}}_1^{\dot{\alpha}}}\,,
\label{S1Spinor}
\end{align}
where $\lambda^\alpha \equiv |i^{+}\rangle^\alpha$ and
 $\tilde\lambda^{\dot\alpha} \equiv |i^{-}\rangle^{\dot\alpha}$.
We consider the explicit action of $S_{n}^{(1)\lambda}$ in Eq.~\eqref{S1Spinor} and $S_{n}^{(1)}$ in Eq.~\eqref{SoftGluonOperators} on $\pol_{1}^{\pm\,\rho}(k_{r_{1}})$ to show equivalence after contraction with the polarization-stripped amplitude. The action on $\pol_{n-1}^{\pm\,\rho}(k_{r_{n-1}})$ follows similarly. We act with \eqn{S1Spinor} on the vectors in Eq.~(\ref{PolSpinor})---with $i\rightarrow1$ and $k_{r}\rightarrow k_{r_{1}}$---in turn:
\begin{align}
S_{n}^{(1)\lambda}\,\pol_{1}^{+\,\rho}(k_{r_{1}})
=-\frac{1}{\spa{1}.{n}}
\frac{\left<r_{1}\right|\gamma^{\rho}\left|n\right]}{\sqrt{2}\spa{r_{1}}.{1}}
=-\frac{\spa{r_{1}}.{n}}{\spa{r_{1}}.{1}\spa{1}.{n}}
\pol_{n}^{+\,\rho}(k_{r_{1}})\,,
\end{align}
and
\begin{align}
S_{n}^{(1)\lambda}\,\varepsilon_{1}^{-\,\rho}(k_{r_{1}})
&=-\frac{1}{\spa{1}.{n}}
\left(
-\frac{\left<1\right|\gamma^{\rho}\left|r_{1}\right]}{\sqrt{2}}
\right)
\left(
-\frac{\spb{r_{1}}.{n}}{\spb{r_{1}}.{1}^{2}}
\right)
\nonumber \\[.2cm]
&=\frac{\spb{r_{1}}.{n}}{\spb{r_{1}}.{1}\spa{1}.{n}}
\varepsilon_{1}^{-\,\rho}(k_{r_{1}})
\nonumber \\[.2cm]
&=\frac{\spb{r_{1}}.{n}}{\spb{r_{1}}.{1}\spa{1}.{n}}
\left[
\varepsilon_{1}^{-\,\rho}(k_{n})
+\frac{\sqrt{2}\spb{r_{1}}.{n}}{\spb{r_{1}}.{1}\spb{n}.{1}}\,k_{1}^{\rho}
\right]
\nonumber \\[.2cm]
&=\frac{\spb{r_{1}}.{n}}{\spb{r_{1}}.{1}\spb{1}.{n}}
\left[
\varepsilon_{n}^{+\,\rho}(k_{1})
-\frac{\sqrt{2}\spb{r_{1}}.{n}}{\spb{r_{1}}.{1}\spa{1}.{n}}\,k_{1}^{\rho}
\right],
\label{S1Lm}
\end{align}
where we used
\begin{align}
\varepsilon_{i}^{-\rho}(k_{r})
=\varepsilon_{i}^{-\rho}(k_{\tilde{r}})
+\frac{\sqrt{2}\spb{r}.{\tilde{r}}}{\spb{r}.{i}\spb{\tilde{r}}.{i}}\,k_{i}^{\rho}\,,
\end{align}
in the second-to-last line.
The last line of Eq. (\ref{S1Lm}) follows from
\begin{align}
\varepsilon_{j}^{+\,\rho}(k_{i})=\frac{\spb{i}.{j}}{\spa{i}.{j}}\,\varepsilon_{i}^{-\,\rho}(k_{j}) \,.
\end{align}
We can write \eqn{S1Lm} more simply as
\begin{align}
S_{n}^{(1)\lambda}\,\varepsilon_{1}^{-\,\rho}(k_{r_{1}})
\cong
\frac{\spb{r_{1}}.{n}}{\spb{r_{1}}.{1}\spb{1}.{n}}\,
\varepsilon_{n}^{+\,\rho}(k_{1})\,,
\end{align}
where the symbol $\cong$ denotes equivalence up to a term proportional
to $k_{1}^{\rho}$.  Such terms will vanish when contracted with the
polarization-stripped $(n-1)$-point amplitude, so we are free to drop
them.  Similar spinor-helicity algebra reveals that the action of
$S^{(1)}_{n}$ from Eq.~\eqref{SoftGluonOperators} on
$\pol_{1}^{\pm\,\rho}(k_{r_{1}})$ yields
\begin{align}
S^{(1)}_{n}\,\varepsilon_{1}^{+\,\rho}(k_{r_{1}})
&=-i\varepsilon_{n\,\mu}^{+}(k_{r_{n}})k_{n\,\sigma}\frac{\S_1^{\mu\sigma}}{\sqrt{2}\,(k_{1}\cdot k_n)}\varepsilon_{1}^{+\,\rho}(k_{r_{1}}) \notag \\
&=
-\frac{\spa{r_{1}}.{n}}{\spa{r_{1}}.{1}\spa{1}.{n}}\,
\varepsilon_{n}^{+\,\rho}(k_{r_{1}}) \,,
\end{align}
and
\begin{align}
S^{(1)}_{n}\,\varepsilon_{1}^{-\,\rho}(k_{r_{1}})
=
\frac{\spb{r_{1}}.{n}}{\spb{r_{1}}.{1}\spb{1}.{n}}\,
\varepsilon_{n}^{+\,\rho}(k_{1})\,.
\end{align}
We can summarize the action of the operators as
\begin{align}
S_{n}^{(1)\lambda}\,\varepsilon_{1}^{\pm\,\rho}(k_{r_{1}})
\cong
S_{n}^{(1)}\,\varepsilon_{1}^{\pm\,\rho}(k_{r_{1}})
=
-\left(\frac{\varepsilon_{1}^{\pm}(k_{r_{1}})\cdot p_{n}}{\sqrt{2}(p_{1}\cdot p_{n})}\right)
\times
\begin{cases}
\varepsilon_{n}^{+\,\rho}(k_{r_{1}}), &\text{for $+$}\,,\\
\varepsilon_{n}^{+\,\rho}(k_{1}), &\text{for $-$}\,.
\end{cases}
\end{align}
We see that, up to terms proportional to $k_{1}^{\rho}$, the action of
$S_{n}^{(1)\lambda}$ and $S_{n}^{(1)}$ on the polarization vectors
yield completely equivalent expressions as expected.

%%%%%%%%%%%%%%%%
\section{Soft limit of $n$-graviton amplitudes}
\label{PureGravitonSection}

In this section, we generalize what has been done for the case of $n$
gluons to the case of $n$ gravitons. As before, we write the amplitude
as a sum of two pieces: the first contains terms where the soft
graviton is attached to one of the other $n-1$ external gravitons,
giving a contribution divergent as $1/q$ for $q \rightarrow 0$, while
in the second the soft graviton attaches to one of the internal
graviton lines and is of $\Ord(q^0)$ in the same limit.  Leaving the
expression uncontracted with polarization tensors for now, we write
\begin{align}
&M_n^{\mu \nu;\mu_1\nu_1 \cdots\mu_{n-1}\nu_{n-1}}  (q; k_1, \dots, k_{n-1}) \notag \\
&\hspace{2cm}= \sum_{i=1}^{n-1} \frac{1}{k_i\cdot\q} \left[ k_i^{\mu} \eta^{\mu_i \alpha} -
i q_{\rho} (\S_i^{\mu \rho} )^{\mu_i \alpha} \right]  \left[ k_i^{\nu} \eta^{\nu_i \beta} -
i q_{\sigma} (\S_i^{\nu \sigma} )^{\nu_i \beta} \right] \nonumber \\
& \hspace{6cm}\times M_{n-1\hspace{.35cm}\alpha\beta}^{\mu_1 \nu_1 \cdots\hphantom{\alpha\beta}\cdots \mu_{n-1} \nu_{n-1} } (k_1, \dots, k_i+q, \dots, k_{n-1} ) \notag \\
&\hspace{2.5cm}+ N_n^{\mu \nu; \mu_1 \nu_1 \cdots \mu_{n-1} \nu_{n-1} }  (q; k_1, \dots, k_{n-1})\,,
\label{EMME}
\end{align}
where
\begin{align}
(\S_i^{\mu \rho})^{\mu_i \alpha} \equiv i\left(\eta^{\mu \mu_i} \eta^{\alpha \rho} - \eta^{\mu \alpha} \eta^{\mu_i \rho}\right) \,.
\label{SSS}
\end{align}
The simple form of the three vertex used in \eqn{EMME} can be obtained
from the standard de Donder gauge one, using current conservation and
tracelessness properties of external polarization tensors and
$M_{n-1}$, as well as assigning terms to $N_n$ where the
$i/k_i \cdot q$ propagator cancels.  We note that it is
important to keep the lowered indices of $M_{n-1}$ in their
appropriate slots.  On-shell gauge invariance implies
\begin{align}
0&= q_{\mu} M_n^{\mu \nu;\mu_1\nu_1 \cdots\mu_{n-1}\nu_{n-1}}  (q; k_1, \dots, k_{n-1}) \notag \\
&=\sum_{i=1}^{n-1}
   \left[ k_i^{\nu} \eta^{\nu_i \beta} - i q_{\rho} (\S_i^{\nu \rho})^{\nu_i \beta} 
  \right] M_{n-1\hspace{.62cm}\beta}^{\mu_1 \nu_1 \cdots  \mu_i \hphantom{\beta} \cdots  \mu_{n-1} \nu_{n-1}} (k_1, \dots, k_i +q, \dots, k_{n-1} ) \nonumber \\
&\hspace{.7cm}+ q_{\mu}  N_n^{\mu \nu ; \mu_1 \nu_1 \cdots \mu_{n-1} \nu_{n-1} }  (q; k_1,  \dots, k_{n-1})\,,
\label{ginv}
\end{align}
provided that as usual we will contract all free indices of $M_n$ with
polarization tensors at the end.  This includes contracting the $\nu$ index
with a polarization vector $\pol_n^\mu$ satisfying $\pol_n \cdot q = 0$.
 Expanding the previous expression
for small $q$, we find that the leading term vanishes because of
momentum conservation, while the next-to-leading term gives two
conditions by taking the symmetric and antisymmetric parts:
\begin{align}
&\frac{1}{2}\sum_{i=1}^{n-1}  \eta^{\mu_i \alpha} \eta^{\nu_i \beta} \left(k_i^{\mu} \frac{\partial}{\partial k_{i\nu} } +  k_i^{\nu} \frac{\partial}{\partial k_{i\mu} }\right) M_{n-1\hspace{.35cm}\alpha\beta}^{\mu_1 \nu_1 \cdots \hphantom{\alpha \beta} \cdots  \mu_{n-1} \nu_{n-1}}  (k_1, \dots, k_i, \dots,  k_{n-1})  \nonumber \\
&\hspace{7.5cm}= -  N_n^{\mu \nu; \mu_1 \nu_1 \cdots \mu_{n-1} \nu_{n-1} }  (0; k_1,  \dots, k_{n-1})\,,
\label{symmone}
\end{align}
and
\begin{align}
\sum_{i=1}^{n-1} \left[ L_i^{\nu \rho} \eta^{\nu_i \beta} + 2  (\S_i^{\nu \rho})^{\nu_i \beta}   \right] M_{n-1\hspace{.62cm}\beta}^{ \mu_1 \nu_1 \cdots  \mu_i \hphantom{\beta}\cdots  \mu_{n-1} \nu_{n-1}}  
(k_1, \dots, k_i, \dots,  k_{n-1})    =0\,.
\label{conseangmom}
\end{align}
As in the earlier cases, we can ignore potential terms that are local in $q$ and vanish when dotted into $q^{\mu}$ since they will not contribute to the desired order.  The first condition determines $N_n^{\mu \nu; \mu_1 \nu_1 \cdots \mu_{n-1} \nu_{n-1} }(0;k_1, \dots, k_{n-1})$ in terms of the amplitude without the soft graviton, while the second one reflects conservation of total angular momentum.  The factor of $2$ in front of the spin term in Eq.~\eqref{conseangmom} reflects the fact that the graviton has spin $2$.

Finally, the terms of order $q^2$ in Eq.~\eqref{ginv} imply the
following condition:
\begin{align}
& \sum_{i=1}^{n-1} q_{\rho} \left[    k_i^{\nu} \eta^{\nu_i \beta} 
 \frac{ \partial^2}{\partial k_{i\rho} \partial k_{i\mu}}
-2i( \S_i^{\nu \rho} )^{\nu_i \beta}  \frac{\partial}{\partial k_{i\mu } }   \right]M_{n-1\hspace{.62cm}\beta}^{\mu_1 \nu_1\cdots  \mu_i \hphantom{\beta} \cdots  \mu_{n-1} \nu_{n-1}}  (k_1, \dots, k_i, \dots,  k_{n-1})  \nonumber \\
& \hspace{2.6cm}= -q_{\rho} \left[ \frac{\partial N_n^{\mu \nu ; \mu_1 \nu_1 \cdots \mu_{n-1} \nu_{n-1} } }{\partial q_\rho}   + \frac{\partial N_n^{\rho \nu ; \mu_1 \nu_1 \cdots \mu_{n-1} \nu_{n-1} }}{\partial q_{\mu}} \right] (0; k_1, \dots, k_{n-1})\,.
\label{2orders}
\end{align}
Using the previous results, for a soft graviton of momentum $q$,
we have
\begin{align}
& M_n^{\mu \nu;\mu_1\nu_1 \cdots\mu_{n-1}\nu_{n-1}}  (q; k_1, \dots, k_{n-1})  \notag \\
&\hspace{.5cm}= \sum_{i=1}^{n-1} \frac{1}{k_i \cdot q} \bigg\{ k_i^{\mu} k_i^{\nu}\eta^{\mu_i\alpha}\eta^{\nu_i\beta} \nonumber \\
&\hspace{1.3cm}- \frac{i}{2}   q_{\rho} \Bigl[k_i^{\mu}\eta^{\mu_i\alpha} \left[ L_i^{\nu \rho} \eta^{\nu_i \beta} + 2 ( \S_i^{\nu \rho})^{\nu_i \beta} \right] +    k_i^{\nu}\eta^{\nu_i\beta}\left[ L_i^{\mu \rho} \eta^{\mu_i \alpha} + 2 ( \S_i^{\mu \rho})^{\mu_i \alpha} \right] \Bigr] \notag \\
&\hspace{1.3cm}-\frac{i}{2}   q_{\rho} q_{\sigma}   \biggl[ k_i^{\nu}\eta^{\mu_i\alpha}\eta^{\nu_i\beta} L_i^{\mu \rho}   \frac{\partial}{ \partial k_{i\sigma}}    - 2i( \S_i^{\mu \rho } )^{\mu_i \alpha}  (\S_i^{\nu \sigma })^{\nu_i \beta}   -  2k_i^{\sigma}\eta^{\nu_i\beta}  (\S_i^{\nu \rho})^{\nu_i \beta } \frac{ \partial}{\partial k_{i\mu}}    \nonumber \\
&\hspace{6.4cm} + 2  \left[\eta^{\mu_i \alpha} k_i^{\mu}  (\S_i^{\nu \rho})^{\nu_i \beta}    + 
\eta^{\nu_i \beta} k_i^{\nu}  (\S_i^{\mu \rho})^{\mu_i \alpha} \right]   \frac{\partial}{\partial k_{i \sigma  }}\biggr]\bigg\} \notag \\
&\hspace{3.5cm}\times M_{n-1\hspace{.35cm}\alpha\beta}^{\mu_1 \nu_1 \cdots\hphantom{\alpha\beta}\cdots \mu_{n-1} \nu_{n-1} } (k_1, \dots, k_i, \dots, k_{n-1} )  \nonumber \\
&\hspace{1cm}  + \frac{1}{2}q_{\rho} \left[ \frac{\partial N_n^{\mu \nu;\mu_1\nu_1 \cdots\mu_{n-1}\nu_{n-1} }}{\partial q_{\rho}}   -  \frac{\partial N_n^{\rho \nu;\mu_1\nu_1 \cdots\mu_{n-1}\nu_{n-1} }}{\partial q_{\mu}}   \right](0;k_1,\dots,k_{n-1}) \notag \\
&\hspace{1cm}+\Ord(q^2)\,.
 \label{amplitot2}
\end{align}

As in the case of gluon scattering, it may seem that we cannot
determine the order $q$ contributions in terms of $M_{n-1}$ because the
antisymmetric part of the matrix $N_n$ is still present in
Eq.~(\ref{amplitot2}).  However, it turns out that there is additional
information from on-shell gauge invariance. When we saturate it with
$q_{\mu}$, we get of course zero because this is the way that
\eqn{amplitot2} is constructed. When we saturate it with
$q_{\nu}$, however, we obtain the extra condition:
\begin{align}
0&= q_{\nu} M_n^{\mu \nu ; \mu_1 \nu_1 \cdots \mu_{n-1} \nu_{n-1}}(q;k_1,\dots,k_{n-1}) \notag \\
&  = q_{\rho} q_{\sigma} 
\Bigg\{ \sum_{i=1}^{n-1}\left[ L_i^{\mu \rho} \eta^{\mu_i \alpha}  + 2 ( \S_i^{\mu \rho})^{\mu_i \alpha} \right] \frac{\partial}{\partial k_{i \sigma}} M_{n-1\hspace{.35cm}\alpha}^{\mu_1 \nu_1 \cdots\hphantom{\alpha}\nu_i\cdots \mu_{n-1} \nu_{n-1} } (k_1, \dots, k_i, \dots, k_{n-1} ) \nonumber \\ 
&\hspace{2.6cm} +i\left[ \frac{\partial N_n^{\mu \sigma;\mu_1\nu_1 \cdots\mu_{n-1}\nu_{n-1} }}{\partial q_{\rho}} -  \frac{\partial N_n^{\rho \sigma;\mu_1\nu_1  \cdots\mu_{n-1}\nu_{n-1} } }{\partial q_{\mu}}  \right] (0;k_1, \dots, k_{n-1} )  \Bigg\}\,,
\label{qnu1}
\end{align}
which implies
\begin{align}
&
 \sum_{i=1}^{n-1}q_{\rho}\left[ L_i^{\mu \rho} \eta^{\mu_i \alpha}  + 2 ( \S_i^{\mu \rho})^{\mu_i \alpha} \right] \frac{\partial}{\partial k_{i \sigma}} M_{n-1\hspace{.35cm}\alpha}^{\mu_1 \nu_1 \cdots\hphantom{\alpha}\nu_i\cdots \mu_{n-1} \nu_{n-1} } (k_1, \dots, k_i, \dots, k_{n-1} ) \nonumber \\ 
&\hspace{2.4cm} =- iq_{\rho}\left[ \frac{\partial N_n^{\mu \sigma;\mu_1\nu_1 \cdots\mu_{n-1}\nu_{n-1} }}{\partial q_{\rho}} -  \frac{\partial N_n^{\rho \sigma;\mu_1\nu_1  \cdots\mu_{n-1}\nu_{n-1} } }{\partial q_{\mu}}  \right] (0;k_1, \dots, k_{n-1} )  \,.
\label{qnu2}
\end{align}
We can now use it in Eq.~(\ref{amplitot2}) to obtain our final expression
giving the soft limit entirely in terms of the $(n-1)$-point amplitude:
\begin{align}
&M_n^{\mu \nu ; \mu_1 \nu_1 \cdots \mu_{n-1} \nu_{n-1}}(q;k_1,\dots,k_{n-1}) \notag \\
&\hspace{.5cm}= \sum_{i=1}^{n-1} \frac{1}{k_i \cdot q} \bigg\{   k_i^{\mu} k_i^{\nu}\eta^{\mu_i\alpha}\eta^{\nu_i\beta}  \nonumber \\
&\hspace{1.3cm}-   \frac{i}{2} q_{\rho} \Bigl[k_i^{\mu}\eta^{\mu_i\alpha} \left[ L_i^{\nu \rho} \eta^{\nu_i \beta} + 2 ( \S_i^{\nu \rho})^{\nu_i \beta} \right] +    k_i^{\nu}\eta^{\nu_i\beta}\left[ L_i^{\mu \rho} \eta^{\mu_i \alpha} + 2 ( \S_i^{\mu \rho})^{\mu_i \alpha} \right] \Bigr] \notag \\
&\hspace{1.3cm}- \frac{1}{2}  q_{\rho}q_{\sigma}  \Bigl[\left[L_i^{\mu \rho} \eta^{\mu_i \alpha}  +  2 (\S_i^{\mu \rho})^{\mu_i \alpha} \right] \left[ L_i^{\nu \sigma} \eta^{\nu_i \beta} + 2(\S_i^{\nu \sigma})^{\nu_i \beta} \right] -  2 (\S_i^{\mu \rho})^{\mu_i \alpha} (\S_i^{\nu \sigma})^{\nu_i \beta}\Bigr]    \bigg\}  \notag \\
&\hspace{2.8cm}\times M_{n-1\hspace{.35cm}\alpha\beta}^{\mu_1 \nu_1 \cdots\hphantom{\alpha\beta}\cdots \mu_{n-1} \nu_{n-1} } (k_1, \dots, k_i, \dots, k_{n-1} ) + \Ord (q^2 )\,.
\label{FINALRESULT}
\end{align}
In order to write our expression in terms of amplitudes, we saturate
with graviton polarization tensors using
$\pol_{\mu\nu}\rightarrow\pol_{\mu}\pol_{\nu}$ where $\pol_{\mu}$ are
spin-one polarization vectors.  As we did for the case with gluons, we
must pass the polarization vectors through the spin-one operators.  We
are then left with
\begin{align}
M_n(q;k_1,\dots,k_{n-1})=\left[S_n^{(0)}+S_n^{(1)}+S_n^{(2)}\right]M_{n-1}(k_1,\dots,k_{n-1})+\Ord(q^2)\,,
\label{SoftGravitonResult}
\end{align}
where
\begin{align}
&S_n^{(0)}\equiv \sum_{i=1}^{n-1}\frac{\pol_{\mu \nu}k_i^{\mu} k_i^{\nu}}{k_i\cdot\q}\,, \notag \\
&S_n^{(1)}\equiv-i\sum_{i=1}^{n-1}\frac{\pol_{\mu \nu}k_i^{\mu}q_{\rho}J_i^{\nu \rho}}{k_i\cdot\q}\,, \notag \\
&S_n^{(2)}\equiv-\frac{1}{2}\sum_{i=1}^{n-1}\frac{\pol_{\mu \nu}q_{\rho} J_i^{\mu \rho} q_{\sigma}  J_i^{\nu \sigma}}{k_i\cdot\q}\,.
\end{align}
Here
\begin{align}
J_i^{\mu\sigma}\equiv L_i^{\mu\sigma}+\Sigma_i^{\mu\sigma}\,,
\end{align}
with
\begin{align}
L_i^{\mu\sigma}
\equiv
i\left(k_i^{\mu} \frac{\partial}{\partial k_{i \sigma}} - 
k_i^{\sigma} \frac{\partial}{\partial k_{i \mu}}\right)\,, \hspace{1.5cm}\Sigma_i^{\mu\sigma}\equiv i\left(\pol_i^{\mu}\frac{\partial}{\partial\pol_{i\sigma}}-\pol_i^{\sigma}\frac{\partial}{\partial\pol_{i\mu}}\right)\,.
\label{eq:spinTerm}
\end{align}
Since the graviton polarization tensor is quadratic in spin-one polarization vectors $\varepsilon_i^{\mu}$, the differential operator in Eq.~\eqref{eq:spinTerm} picks up factors of 2 as required for Eq.~\eqref{SoftGravitonResult} to be compatible with Eq.~\eqref{FINALRESULT}.

In conclusion, in the case of a soft graviton, on-shell gauge
invariance completely determines the first two subleading
contributions. Using the Kawai-Lewellen-Tye relations~\cite{KLT} we have
generated graviton amplitudes with  formal polarization tensors
up to six points.  Using these we analytically confirmed
\eqn{SoftGravitonResult} through five points and numerically through
six points.

%%%%%%%%%%%%%%%%%%%%%%%%%%%%%%%%%%%%%%%%%%%%%%%%%%%%%%%%%%%%
\section{Comments on Loop Corrections}
\label{LoopSection}

In gauge and gravity theories in four dimensions, the operators
describing the soft behavior have nontrivial loop
corrections~\cite{LoopCorrections,HeHuang}.  Indeed, in QCD loop
corrections linked to infrared singularities are present already at
leading order in the soft
limit~\cite{OneLoopSoftBern,OneLoopSoftKosower}.  One may wonder how
loop corrections to the soft operators arise from the perspective of
the constraints imposed by on-shell gauge invariance.  In this section
we explain this.  We first describe the case of gauge theory before
turning to gravity.

%%%%%%%%% FIGURE %%%%%%%%%%%%%%%
\begin{figure}[ht]
\centering
\includegraphics[scale=.4]{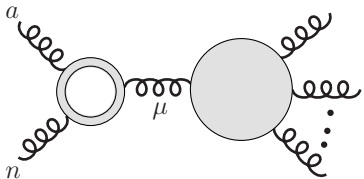}
\caption[a]{The potential factorizing contributions to 
the one-loop corrections to the leading soft function which then cancel. Leg
$n$ is the soft leg which carries momentum $q$. At subleading order
there are additional contributions.
}
\label{FactorizingLoopYMFigure}
\end{figure}
%%%%%%%%%%%%%%%%%%%%%%%%%%%%%%%%

As explained in Ref.~\cite{OneLoopSoftBern}, we can separate the
contributions into two distinct sources.  The first source of
potential corrections is the ``factorizing'' one that, for leading
order, arises from loop corrections of the form displayed in
\fig{FactorizingLoopYMFigure}~\cite{OneLoopSoftBern,LoopCorrections,HeHuang}.
The second type of contributions are ``nonfactorizing''
infrared-divergent pieces that can come from discontinuities in the
amplitudes associated with infrared divergences~\cite{BernChalmers}.
(Alternatively these nonfactorizing contributions can be pushed into
factorizing contributions that have light-cone denominators coming
from a careful application of unitarity~\cite{OneLoopSoftKosower}.)

Here we will focus on the factorizing pieces. In gauge theory we will
explain why they do not enter in the leading soft
behavior~\cite{OneLoopSoftBern,OneLoopSoftKosower}.  For the case of
scalars in the loops, which is an especially clean case since there
are no infrared singularities even for massless scalars, we show that
there are no factorizing loop corrections at the leading and first subleading
orders of the soft-graviton expansion. This suppression was noticed
earlier in explicit examples of soft limits of one-loop
infrared-finite gravity amplitudes~\cite{LoopCorrections,HeHuang}.

\subsection{Gauge Theory}
\label{YMLoopSubSection}

%%%%%%%%% FIGURE %%%%%%%%%%%%%%%
\begin{figure}[ht]
\centering
\includegraphics[scale=.4]{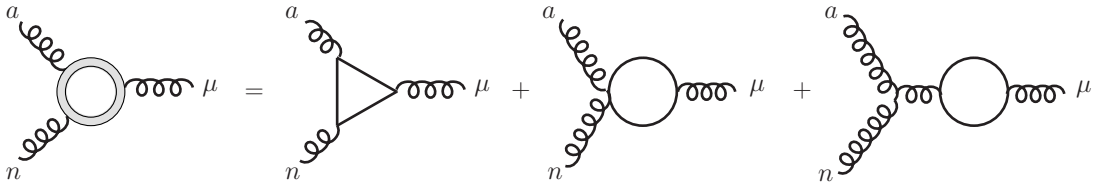}
\caption[a]{The diagrams with potential
  factorizing contributions to the one-loop soft function. At
  subleading order there are additional contributions.  }
\label{DmuDiagsYMFigure}
\end{figure}
%%%%%%%%%%%%%%%%%%%%%%%%%%%%%%%%

As a warm up to the gravity case, we first discuss the well-understood
gauge-theory case.  The explicit forms of the factorizing one-loop
corrections to the soft behavior have been described in some detail in
Refs.~\cite{OneLoopSoftBern,OneLoopSoftKosower} for QCD at leading
order in the soft (and collinear) limits.

For the case of external gluons, the potential factorizing
contributions to one-loop modifications of the leading soft behavior are shown
in \fig{FactorizingLoopYMFigure}.  We can expand these corrections
into triangle and bubble diagrams as shown in
\fig{DmuDiagsYMFigure}. As derived in Ref.~\cite{OneLoopSoftBern},
these diagrams evaluate to
\vspace{.1cm}
\begin{align}
D^{\mu, \rm fact} = \frac{i}{\sqrt{2}} \frac{1}{3} \frac{1}{(4 \pi)^2}
\Bigl(1 - \frac{n_{\! f}}{N_c} + \frac{n_s}{N_c} \Bigr)
(q - k_a)^\mu 
 \Bigl[ (\pol_n \cdot \pol_a) - \frac{(q \cdot \pol_a) (k_a \cdot \pol_n)}
                                 {(k_a \cdot q)} \Bigr] \,,
\label{Dmu}
\end{align}
\vspace{.1cm}%
where $n_{\!f}$ is the number of fundamental representation fermions,
$n_s$ the number of fundamental representation complex scalars (using
the normalization conventions of Ref.~\cite{OneLoopSoftBern}), and
$N_c$ is the number of colors.  As usual we take the soft momentum of
leg $n$ to be $q$.  After integration this result is both ultraviolet-
and infrared-finite, so we have taken $\eps = 0$ in the final
integrated result.  The all orders in $\eps$ form of \eqn{Dmu} is
given in Refs.~\cite{OneLoopSoftBern, OneLoopSoftKosower}. 

The result (\ref{Dmu}) has a few surprising features that explain 
why we cannot use it to obtain the full subleading soft correction via gauge 
invariance.  The first is that the correction
to the three-vertex is nonlocal because of the pole in $q \cdot k_a$
that arises from the loop integration. Indeed, after we include the
intermediate propagator $-i/(k_a+q)^2$, there is a double
pole\footnote{While this might seem to violate basic factorization
  properties of field theories, in fact it does not, because for real
  momenta the double pole is reduced to a single pole.  See
  Refs.~\cite{NontrivialFactorization1,NontrivialFactorization2} for a detailed discussion of
  this phenomenon.}  in $q \cdot k_a$.  A second curious feature is
that \eqn{Dmu} is gauge invariant by itself: It
vanishes when $\pol_n^{\mu}$ is replaced by $q^{\mu}\equiv k_n^{\mu}$
for any value of the intermediate off-shell momentum.  The nonlocal
nature of the result is what allows us to write such a gauge-invariant
term with the correct dimensions.

A third feature is that, in fact, there is no contribution from
\eqn{Dmu} to the leading one-loop correction to the soft function, as
noted in Refs.~\cite{OneLoopSoftBern,OneLoopSoftKosower}.  To see this,
we sew \eqn{Dmu} onto the rest of the amplitude across
the factorization channel:
\begin{align}
D_{\mu}^{\rm fact} \frac{-i}{2q\cdot k_a} {\cal J}^\mu \,,
\end{align}
as illustrated in \Fig{FactorizingLoopYMFigure}.
We observe that ${\cal J}^\mu$ is a conserved current:
\begin{equation}
(q + k_a)_\mu {\cal J}^\mu = 0\,,
\end{equation}
assuming that all the remaining legs are contracted with on-shell
polarizations.  This immediately implies
\begin{equation}
  D_{\mu}^{\rm fact} \frac{-i}{2q \cdot k_a} {\cal J}^\mu = \Ord(q^0)\,,
\label{VanishingLeading}
\end{equation}
because $D_\mu^{\rm fact}$ is proportional to $(q - k_a)_\mu$ which
is equivalent to $2q_\mu$ when dotted into a conserved current.  This
reproduces the fact that there is no leading $\Ord(1/q)$ factorizing
contribution to the one-loop soft
function~\cite{OneLoopSoftBern,OneLoopSoftKosower}.

Unfortunately, the $\Ord(q^0)$ terms in the full factorizing
corrections are not under control via gauge invariance.  Once we
allow for an extra $1/(q\cdot k_a)$ nonlocality arising from the loop
integration, we lose control over the subleading piece.  This cannot
happen at tree level because there is no source of a second factor of
$1/(q\cdot k_a)$. In fact, \eqn{Dmu} is incomplete for capturing all
subleading contributions.  The additional contributions have already
been described in some detail at one loop on a case-by-case basis in
Refs.~\cite{NontrivialFactorization1, NontrivialFactorization2,
  DunbarFactorization}.  Unfortunately, no universal factorization
formula is known for these types of corrections, although case-by-case
their forms appear to be relatively simple.  An example of this type
of nontrivial factorization can be found in Eq.~(61) of
Ref.~\cite{NontrivialFactorization1} or Eq.~(3.9) of
Ref.~\cite{HeHuang}; the precise form of the correction depends on the
helicities of other legs.

Interestingly, these contributions resemble an anomaly that seemingly
vanishes if we take the loop integrand strictly in four dimensions.
This arises from the integration where a $1/\eps$ ultraviolet pole
cancels a factor of $\eps$ from numerator algebra, leaving terms of
$\Ord(1)$.  This is reminiscent of the way the chiral anomaly
arises from triangle diagrams in dimensional regularization.  Indeed,
for the single minus-helicity case discussed in
Refs.~\cite{LoopCorrections, HeHuang}, not only does this contribution
vanish but the entire amplitude would vanish if we were not careful to
keep in the integrand in $D=4-2\eps$ instead of four dimensions.  It
is interesting that these types of contributions do not appear in
supersymmetric theories.

Besides the loop contributions described above, there is a second type
of loop correction to the soft operators (\ref{SoftGluonOperators})
arising from non-smoothness in the amplitude due to infrared
singularities~\cite{BernChalmers}.  In QED the integrals are smooth
because the electron mass acts as an infrared cutoff, but in QCD or
gravity there is no such physical cutoff on gluons or gravitons.  It
is therefore much more difficult to consistently introduce a mass regulator
without breaking gauge symmetry or altering the number of propagating
degrees of freedom.  As is standard practice, it is far simpler to use
dimensional regularization.  As discussed in some detail in
Refs.~\cite{BernChalmers, OneLoopSoftBern, LoopCorrections}, as gluons
become soft or collinear, the matrix elements develop discontinuities
that are absorbed into modifications of the loop splitting or soft
operators.  Alternatively, by using light-cone gauge or carefully
applying unitarity, one introduces light-cone denominators containing
a reference momentum, and one can push all contributions into factorizing
diagrams~\cite{StermanSoft,OneLoopSoftKosower}.  Either way, the
conclusion is the same: There are nontrivial contributions due to
infrared singularities not accounted for in the naive tree-level
soft limit.

\subsection{Gravity}

%%%%%%%%% FIGURE %%%%%%%%%%%%%%%
\begin{figure}[ht]
\centering
\includegraphics[scale=.4]{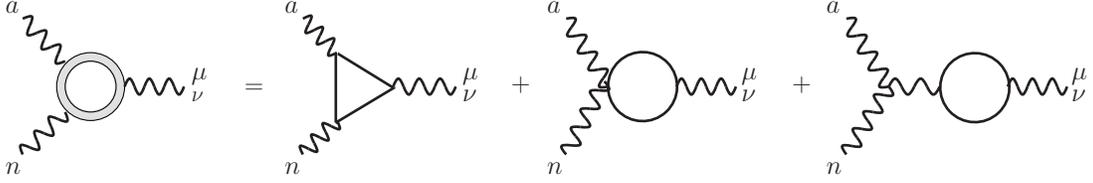}
\caption[a]{The diagrams with potential factorizing contributions to
  the one-loop soft behavior in gravity with a scalar in the
  loop. This captures all such potential leading and first subleading 
  contributions, but it is incomplete at second subleading order.
}
\label{DmuDiagsFigure}
\end{figure}
%%%%%%%%%%%%%%%%%%%%%%%%%%%%%%%%

We now show that the situation in gravity is similar.  Here the
dimensionful coupling ensures that there are no loop corrections at leading
order~\cite{OneLoopMHVGrav}, only one-loop corrections at the first
subleading order, and only up to two-loop corrections at second
subleading order~\cite{LoopCorrections}.  Thus, we need only analyze
one loop to show that the factorizing contributions do not modify the
soft operator at first subleading order.

We focus on the case of a scalar in the loop.  This case is
particularly transparent because there are no infrared singularities
associated with scalars circulating in a loop~\cite{GravityIR}. This
allows us to study the soft behavior without being entangled with the
issue of infrared divergences. We can determine the behavior through
the first subleading soft order simply by computing the diagrams in
\fig{DmuDiagsFigure}.  

We have carried out the analogous computation to the one performed in
Ref.~\cite{OneLoopSoftBern} for gluons, but for gravity with a real
scalar in the loop.  The result of this computation is
\begin{align}
\mathcal{D}^{\mu\nu,\mathrm{fact,s}}&=
-\frac{i}{(4\pi)^2}\left(\frac{\kappa}{2}\right)^3\frac{1}{30q\cdot k_a}\Big((\pol_n\cdot\pol_a)(q\cdot k_a)
-(q\cdot\pol_a)(k_a\cdot\pol_n)\Big)^2k_a^{\mu}k_a^{\nu} + \Ord(q^2) \,,
\label{ScalarLoopDiagrams}
\end{align}
where we have kept all terms involving no more than one overall power
of the soft momentum $q\equiv k_n$.  Such terms naively appear to contribute at the first subleading order in the correction to the amplitude.  However, as in the gauge-theory case, the diagrams
$\mathcal{D}^{\mu\nu,\mathrm{fact,s}}$ contract into a current $\cal{J}_{\mu\nu}$ which results in a suppression of an extra factor of the soft momentum $q$.  In the gravity case we find
\begin{align}
(k_a +q)^\mu {\cal J}_{\mu \nu} = f(k_i,\varepsilon_i)(k_a+q)_{\nu}\,,
\end{align}
where $f$ is some function of the momenta and polarizations of both the hard and soft legs.  With $k_a^{\mu}k_a^{\nu}$ contracting with $\cal{J}_{\mu\nu}$, we then have
\begin{align}
k_a^{\mu}k_a^{\nu}\cal{J}_{\mu\nu}&=(k_a+q)^{\mu}(k_a+q)^{\nu}\mathcal{J}_{\mu\nu}+\Ord (q) \notag \\
&=f(k_i,\varepsilon_i)(k_a+q)^2+\Ord(q) \notag \\
&=2f(k_i,\varepsilon_i)q\cdot k_a+\Ord(q) \notag \\
&=\Ord(q)\,.
\end{align}
Therefore as far the correction to the amplitude is concerned, we can effectively view $\mathcal{D}^{\mu\nu,\mathrm{fact,s}}$ as being of order $q^2$.  We then finally have
\begin{align}
\mathcal{D}^{\mu\nu,\mathrm{fact,s}} \frac{i}{2 q\cdot k_a}
  {\cal J}_{\mu \nu} = \Ord(q)\,.
\label{ScalarLoopBehavior}
\end{align}
After including the $1/q$ from the
intermediate propagator, we find the potential correction to the soft
operator is of $\Ord(q)$ and therefore does not modify the first
subleading soft behavior.  Unfortunately, for the second subleading
soft behavior we lose control, in much the same way that we did for the first
subleading behavior of gauge theory.  Indeed, nontrivial contributions
are found in explicit examples~\cite{LoopCorrections,HeHuang}.

As in the QCD case (\ref{Dmu}), we expect the cases with other
particles circulating in the loop to be similar and that factorizing
contributions not linked to infrared singularities should appear
starting only at the second subleading order in the soft expansion.
In addition, the explicit gravity examples studied in
Refs.~\cite{LoopCorrections, HeHuang} are exactly in line with this
expectation.  We leave a discussion of cases with infrared
singularities to future work.

\section{Conclusions}
\label{ConclusionSection}

In this paper we extended Low's proof of the universality of
subleading behavior of photons to nonabelian gauge theory and to
gravity. In particular, we showed that in gauge theory, on-shell gauge
invariance can be used to fully determine the first subleading soft-gluon
behavior at tree level.  In gravity the first two subleading terms in
the soft expansion found in Ref.~\cite{CachazoStrominger} can also be
fully determined from on-shell gauge invariance.  Our discussion is
similar to the ones given by Low~\cite{LowTheorem} for photons and by
Jackiw~\cite{Jackiw} for gravitons coupled to a scalar at four points.
We focused mainly on $n$-gluon and $n$-graviton amplitudes, but also
discussed simpler cases with scalars.

A motivation for studying soft-graviton theorems is to understand
their relation to the extended BMS symmetry.  It will, of course,
be very important to understand how BMS symmetry relates to the proof
of soft properties in $n$-graviton amplitudes given here.

Unlike the case of photons, for gluons there are loop corrections to
the soft operators starting at leading order.  In gauge theory,
leading-order corrections are linked to infrared singularities, while
subleading-order corrections can also arise from contributions not
linked to infrared singularities.  Gravity also has loop corrections
but not at leading order.  In this paper we proved that for the case
of a scalar circulating in the loop, there is no modification to the
soft behavior of graviton amplitudes until the second subleading
order. We expect this to hold in general for contributions not linked
to infrared singularities.  On the other hand, graviton loop
contributions that are infrared divergent give corrections to the soft
operators starting at the first subleading
order~\cite{LoopCorrections}, using the standard definition of
dimensional regularization.  Since infrared singularities
are well-understood, we do not expect this to be too disruptive for
studying the consequences of extended BMS symmetry at loop level.  We
will describe loop level in more detail elsewhere.

\subsection*{Acknowledgments}

We thank Johannes Broedel, Marius de 
Leeuw, Yu-tin Huang, Henrik Johansson, Daniele
Marmiroli, Raffaele Marotta, Jan Plefka, Radu Roiban, Matteo Rosso and Andrew
Strominger for helpful discussions.  This work was supported by the US
Department of Energy under Award Number DE-{S}C0009937. We also
gratefully acknowledge Mani Bhaumik for his generous support.

%%%%%%%%%%%%%%%%%%%%%%%%%%%%%%%%%%%%%%%%%%%%%%%%%%


\begin{thebibliography}{99}

%+% 3 refs
\bibitem{Strominger}
A.~Strominger,
%``On BMS Invariance of Gravitational Scattering,''
arXiv:1312.2229 [hep-th];\\
%%CITATION = ARXIV:1312.2229;%%
%
T.~He, V.~Lysov, P.~Mitra and A.~Strominger,
%``BMS supertranslations and Weinberg's soft graviton theorem,''
arXiv:1401.7026 [hep-th];\\
%%CITATION = ARXIV:1401.7026;%%
%
D.~Kapec, V.~Lysov, S.~Pasterski and A.~Strominger,
%``Semiclassical Virasoro Symmetry of the Quantum Gravity S-Matrix,''
arXiv:1406.3312 [hep-th].
%%CITATION = ARXIV:1406.3312;%%

%+% 6 refs
\bibitem{CachazoStrominger}
 F.~Cachazo and A.~Strominger,
  %``Evidence for a New Soft Graviton Theorem,''
  arXiv:1404.4091 [hep-th].
  %%CITATION = ARXIV:1404.4091;%%

%+% 1 ref
\bibitem{BMS}
H.~Bondi, M.~G.~J.~van der Burg and A.~W.~K.~Metzner,
%``Gravitational waves in general relativity. 7. Waves from axisymmetric isolated systems,''
Proc.\ Roy.\ Soc.\ Lond.\ A {\bf 269}, 21 (1962);\\
%%CITATION = PRSLA,A269,21;%%
%
R.~K.~Sachs,
%``Gravitational waves in general relativity. 8. Waves in asymptotically flat space-times,''
Proc.\ Roy.\ Soc.\ Lond.\ A {\bf 270}, 103 (1962).
%%CITATION = PRSLA,A270,103;%%

%+% 1 ref
\bibitem{ExtendedBMS}
 G.~Barnich and C.~Troessaert,
  %``Symmetries of asymptotically flat 4 dimensional spacetimes at null infinity revisited,''
  Phys.\ Rev.\ Lett.\  {\bf 105}, 111103 (2010)
  [arXiv:0909.2617 [gr-qc]];\\
  %%CITATION = ARXIV:0909.2617;%%
 G.~Barnich and C.~Troessaert,
  %``BMS charge algebra,''
  JHEP {\bf 1112}, 105 (2011)
  [arXiv:1106.0213 [hep-th]];\\
  %%CITATION = ARXIV:1106.0213;%%
G.~Barnich and C.~Troessaert,
  %``Comments on holographic current algebras and asymptotically flat four dimensional spacetimes at null infinity,''
  JHEP {\bf 1311}, 003 (2013)
  [arXiv:1309.0794 [hep-th]].
  %%CITATION = ARXIV:1309.0794;%%

%+% 2 refs
\bibitem{SoftGluonProof}
 E.~Casali,
  %``Soft sub-leading divergences in Yang-Mills amplitudes,''
  arXiv:1404.5551 [hep-th].
  %%CITATION = ARXIV:1404.5551;%%

%+% 1 ref
\bibitem{Volovich}
B.~U.~W.~Schwab and A.~Volovich,
%``Subleading soft theorem in arbitrary dimension from scattering equations,''
 arXiv:1404.7749 [hep-th];\\
 %%CITATION = ARXIV:1404.7749;%%
%
N.~Afkhami-Jeddi,
%``Soft Graviton Theorem in Arbitrary Dimensions,''
arXiv:1405.3533 [hep-th].
%%CITATION = ARXIV:1405.3533;%%

%+% 1 ref
\bibitem{Conformal}
A.~J.~Larkoski,
%``Conformal Invariance of the Subleading Soft Theorem in Gauge Theory,''
arXiv:1405.2346 [hep-th].
%%CITATION = ARXIV:1405.2346;%%

%+% 3 refs
\bibitem{IntegrandSoft}
 M.~Bianchi, S.~He, Y.-t.~Huang and C.~Wen,
  %``More on Soft Theorems: Trees, Loops and Strings,''
  arXiv:1406.5155 [hep-th].
  %%CITATION = ARXIV:1406.5155;%%

%+% 1 ref
\bibitem{TwistorSoft}
T.~Adamo, E.~Casali and D.~Skinner,
%``Perturbative gravity at null infinity,''
arXiv:1405.5122 [hep-th];\\
%%CITATION = ARXIV:1405.5122;%%
%
Y.~Geyer, A.~E.~Lipstein and L.~Mason,
%``Ambitwistor strings at null infinity and subleading soft limits,''
arXiv:1406.1462 [hep-th].
%%CITATION = ARXIV:1406.1462;%%

%+% 1 ref
\bibitem{StringSoft}
B.~U.~W.~Schwab,
%``Subleading Soft Factor for String Disk Amplitudes,''
arXiv:1406.4172 [hep-th].
%%CITATION = ARXIV:1406.4172;%%

%+% 1 ref
\bibitem{LowFourPt}
F.~E.~Low,
%``Scattering of light of very low frequency by systems of spin 1/2,''
Phys.\ Rev.\  {\bf 96}, 1428 (1954);\\
%%CITATION = PHRVA,96,1428;%%
%
M.~Gell-Mann and M.~L.~Goldberger,
%``Scattering of low-energy photons by particles of spin 1/2,''
Phys.\ Rev.\  {\bf 96}, 1433 (1954);\\
%%CITATION = PHRVA,96,1433;%%
%
S.~Saito,
%``Low-energy theorem for Compton scattering,''
Phys.\ Rev.\  {\bf 184}, 1894 (1969).
%%CITATION = PHRVA,184,1894;%%

%+% 5 refs
\bibitem{LowTheorem}
F.~E.~Low,
%``Bremsstrahlung of very low-energy quanta in elementary particle collisions,''
Phys.\ Rev.\  {\bf 110}, 974 (1958).
%%CITATION = PHRVA,110,974;%%

%+% 4 refs
\bibitem{Weinberg}
S.~Weinberg,
%``Photons and Gravitons in s Matrix Theory: Derivation of Charge Conservation and Equality of Gravitational and Inertial Mass,''
Phys.\ Rev.\  {\bf 135}, B1049 (1964);\\
%%CITATION = PHRVA,135,B1049;%%
%
S.~Weinberg,
%``Infrared photons and gravitons,''
Phys.\ Rev.\  {\bf 140}, B516 (1965).
%%CITATION = PHRVA,140,B516;%%

%+% 1 ref
\bibitem{OtherSoftPhotons}
T.~H.~Burnett and N.~M.~Kroll,
  %``Extension of the low soft photon theorem,''
  Phys.\ Rev.\ Lett.\  {\bf 20}, 86 (1968);\\
  %%CITATION = PRLTA,20,86;%%
%
J.~S.~Bell and R.~Van Royen,
  %``On the low-burnett-kroll theorem for soft-photon emission,''
  Nuovo Cim.\ A {\bf 60}, 62 (1969);\\
  %%CITATION = NUCIA,A60,62;%%
%
 V.~Del Duca,
  %``High-energy Bremsstrahlung Theorems for Soft Photons,''
  Nucl.\ Phys.\ B {\bf 345}, 369 (1990).
  %%CITATION = NUPHA,B345,369;%%

%+% 1 ref
\bibitem{DFFR}
 V.~de Alfaro, S.~Fubini, G.~Furlan and C.~Rossetti,
{\it ``Currents in hadron physics''}, 
North-Holland, Amsterdam 1974, see Chapter 3.
 %%CITATION = INSPIRE-940209;%%

%+% 3 refs
\bibitem{OneLoopMHVGrav}
Z.~Bern, L.~J.~Dixon, M.~Perelstein and J.~S.~Rozowsky,
%``Multileg one loop gravity amplitudes from gauge theory,''
Nucl.\ Phys.\ B {\bf 546}, 423 (1999)
[hep-th/9811140].
%%CITATION = HEP-TH/9811140;%%

%+% 2 refs
\bibitem{GrossJackiw}
D.~J.~Gross and R.~Jackiw,
%``Low-Energy Theorem for Graviton Scattering,''
Phys.\ Rev.\  {\bf 166}, 1287 (1968).
%%CITATION = PHRVA,166,1287;%%

%+% 3 refs
\bibitem{Jackiw}
R.~Jackiw,
%``Low-Energy Theorems for Massless Bosons: Photons and Gravitons,''
 Phys.\ Rev.\  {\bf 168}, 1623 (1968).
%%CITATION = PHRVA,168,1623;%%

%+% 1 ref
\bibitem{WhiteYM}
E.~Laenen, G.~Stavenga and C.~D.~White,
%``Path integral approach to eikonal and next-to-eikonal exponentiation,''
JHEP {\bf 0903}, 054 (2009)
[arXiv:0811.2067 [hep-ph]];\\
%%CITATION = ARXIV:0811.2067;%%
%
E.~Laenen, L.~Magnea, G.~Stavenga and C.~D.~White,
%``Next-to-eikonal corrections to soft gluon radiation: a diagrammatic approach,''
JHEP {\bf 1101}, 141 (2011)
[arXiv:1010.1860 [hep-ph]].
%%CITATION = ARXIV:1010.1860;%%

%+% 1 ref
\bibitem{WhiteGrav}
C.~D.~White,
%``Factorization Properties of Soft Graviton Amplitudes,''
JHEP {\bf 1105}, 060 (2011)
[arXiv:1103.2981 [hep-th]].
%%CITATION = ARXIV:1103.2981;%%

%+% 15 refs
\bibitem{LoopCorrections}
Z.~Bern, S.~Davies and J.~Nohle,
%``On Loop Corrections to Subleading Soft Behavior of Gluons and Gravitons,''
arXiv:1405.1015 [hep-th].
%%CITATION = ARXIV:1405.1015;%%

%+% 10 refs
\bibitem{HeHuang}
 S.~He, Y.-t.~Huang and C.~Wen,
  %``Loop Corrections to Soft Theorems in Gauge Theories and Gravity,''
  arXiv:1405.1410 [hep-th].
  %%CITATION = ARXIV:1405.1410;%%

%+% 14 refs
\bibitem{OneLoopSoftBern}
Z.~Bern, V.~Del Duca and C.~R.~Schmidt,
  %``The Infrared behavior of one loop gluon amplitudes at next-to-next-to-leading order,''
  Phys.\ Lett.\ B {\bf 445}, 168 (1998)
  [hep-ph/9810409];\\
  %%CITATION = HEP-PH/9810409;%%
%
Z.~Bern, V.~Del Duca, W.~B.~Kilgore and C.~R.~Schmidt,
  %``The infrared behavior of one loop QCD amplitudes at next-to-next-to leading order,''
  Phys.\ Rev.\ D {\bf 60}, 116001 (1999)
  [hep-ph/9903516].
  %%CITATION = HEP-PH/9903516;%%

%+% 10 refs
\bibitem{OneLoopSoftKosower}
 D.~A.~Kosower and P.~Uwer,
  %``One loop splitting amplitudes in gauge theory,''
  Nucl.\ Phys.\ B {\bf 563}, 477 (1999)
  [hep-ph/9903515];\\
  %%CITATION = HEP-PH/9903515;%%
%
 D.~A.~Kosower,
  %``All orders singular emission in gauge theories,''
  Phys.\ Rev.\ Lett.\  {\bf 91}, 061602 (2003)
  [hep-ph/0301069].
  %%CITATION = HEP-PH/0301069;%%

%+% 2 refs
\bibitem{FreddyNew}
 F.~Cachazo and E.~Y.~Yuan,
  %``Are Soft Theorems Renormalized?,''
  arXiv:1405.3413 [hep-th].
  %%CITATION = ARXIV:1405.3413;%%

%+% 2 refs
\bibitem{GravityIR}
 S.~G.~Naculich and H.~J.~Schnitzer,
  %``Eikonal methods applied to gravitational scattering amplitudes,''
  JHEP {\bf 1105}, 087 (2011)
  [arXiv:1101.1524 [hep-th]];
  %%CITATION = ARXIV:1101.1524;%%
%
 R.~Akhoury, R.~Saotome and G.~Sterman,
  %``Collinear and Soft Divergences in Perturbative Quantum Gravity,''
  Phys.\ Rev.\ D {\bf 84}, 104040 (2011)
  [arXiv:1109.0270 [hep-th]].
  %%CITATION = ARXIV:1109.0270;%%
  
%+% 1 ref
\bibitem{NewPaper}
J.~Broedel, M.~de~Leeuw, J.~Plefka, and M. Rosso,  arXiv:1406.6574 [hep-th].
%%CITATION = ARXIV:1406.6574;%%

%+% 2 refs
\bibitem{DixonReview}
L.~J.~Dixon,
%``Calculating scattering amplitudes efficiently,''
in Boulder 1995, QCD and beyond, p. 539-582
[hep-ph/9601359].
%%CITATION = HEP-PH/9601359;%%

\bibitem{WeinbergBook}
S.~Weinberg, ``The Quantum Theory of Fields. Vol. 1: Foundations,''
Cambridge Univ. Press (1995).

%+% 1 ref
\bibitem{KLT}
 H.~Kawai, D.~C.~Lewellen and S.~H.~H.~Tye,
  %``A Relation Between Tree Amplitudes of Closed and Open Strings,''
  Nucl.\ Phys.\ B {\bf 269}, 1 (1986).
  %%CITATION = NUPHA,B269,1;%%

%+% 3 refs
\bibitem{BernChalmers}
Z.~Bern and G.~Chalmers,
%``Factorization in one loop gauge theory,''
Nucl.\ Phys.\ B {\bf 447}, 465 (1995)
[hep-ph/9503236].
%%CITATION = HEP-PH/9503236;%%

%+% 2 refs
\bibitem{NontrivialFactorization1}
Z.~Bern, L.~J.~Dixon and D.~A.~Kosower,
  %``On-shell recurrence relations for one-loop QCD amplitudes,''
  Phys.\ Rev.\ D {\bf 71}, 105013 (2005)
  [hep-th/0501240].
  %%CITATION = HEP-TH/0501240;%%

\bibitem{NontrivialFactorization2}
  Z.~Bern, L.~J.~Dixon and D.~A.~Kosower,
  %``The last of the finite loop amplitudes in QCD,''
  Phys.\ Rev.\ D {\bf 72}, 125003 (2005)
  [hep-ph/0505055];\\
  %%CITATION = HEP-PH/0505055;%%
%
Z.~Bern, L.~J.~Dixon and D.~A.~Kosower,
  %``Bootstrapping multi-parton loop amplitudes in QCD,''
  Phys.\ Rev.\ D {\bf 73}, 065013 (2006)
  [hep-ph/0507005].
  %%CITATION = HEP-PH/0507005;%%

%+% 1 ref
\bibitem{DunbarFactorization}
D.~Vaman and Y.~-P.~Yao,
%``Constraints and Generalized Gauge Transformations on Tree-Level Gluon and Graviton Amplitudes,''
 JHEP {\bf 1011}, 028 (2010)
[arXiv:1007.3475 [hep-th]];\\
 %%CITATION = ARXIV:1007.3475;%%
%
S.~D.~Alston, D.~C.~Dunbar and W.~B.~Perkins,
%``Complex Factorisation and Recursion for One-Loop Amplitudes,''
Phys.\ Rev.\ D {\bf 86}, 085022 (2012)
[arXiv:1208.0190 [hep-th]].
%%CITATION = ARXIV:1208.0190;%%

%+% 1 ref
\bibitem{StermanSoft}
J.~C.~Collins and G.~F.~Sterman,
%``Soft Partons in {QCD},''
Nucl.\ Phys.\ B {\bf 185}, 172 (1981);\\
%%CITATION = NUPHA,B185,172;%%
%
J.~C.~Collins, D.~E.~Soper and G.~F.~Sterman,
%``Soft Gluons and Factorization,''
Nucl.\ Phys.\ B {\bf 308}, 833 (1988).
%%CITATION = NUPHA,B308,833;%%

\end{thebibliography}
\end{document}